\documentclass{article}

\usepackage{arxiv}

\usepackage[utf8]{inputenc} 
\usepackage[T1]{fontenc}    
\usepackage{hyperref}       
\usepackage{url}            
\usepackage{booktabs}       
\usepackage{amsfonts}       
\usepackage{nicefrac}       
\usepackage{microtype}      
\usepackage{lipsum}		
\usepackage{graphicx}
\usepackage{doi}
\usepackage{authblk}
\usepackage[inkscapelatex=false]{svg}
\usepackage{float}
 \usepackage{amsmath, amsfonts, amssymb}
\usepackage[numbers,square]{natbib}  
\usepackage{titlesec}
\usepackage[labelfont=bf]{caption}  

\title{Learning Reaction-Diffusion Kinetics from Mechanical Information}

\author[a]{Royal C. Ihuaenyi}
\author[b,c]{Hongbo Zhao}
\author[a]{Ruqing Fang}
\author[a]{Ruobing Bai}
\author[d,e]{Martin Z. Bazant}
\author[a,*]{Juner Zhu}

\affil[a]{Department of Mechanical and Industrial Engineering, Northeastern University}
\affil[b]{Department of Physics, University of California, San Diego}
\affil[c]{Department of Chemistry and Biochemistry, University of California, San Diego}
\affil[d]{Department of Chemical Engineering, Massachusetts Institute of Technology}
\affil[e]{Department of Mathematics, Massachusetts Institute of Technology}

\renewcommand{\shorttitle}{Learning Reaction-Diffusion Kinetics from Mechanical Information}

\begin{document}
\maketitle

\begin{abstract}
A central challenge in materials science is characterizing chemical processes that are elusive to direct measurement, particularly in functional materials operating under realistic conditions. Here, we demonstrate that mechanical strain fields contain sufficient information to reconstruct hidden chemical kinetics in coupled chemomechanical systems. Our partial differential equation-constrained learning framework decodes concentration-dependent diffusion kinetics, thermodynamic driving forces, and spatially heterogeneous reaction rates solely from mechanical observations. Using battery electrode materials as a model system, we demonstrate that the framework can accurately identify complex constitutive laws governing three distinct scenarios: classical Fickian diffusion, spinodal decomposition with pattern formation, and heterogeneous electrochemical reactions with spatial rate variations. The approach demonstrates robustness while maintaining accuracy with limited spatial data and reasonable experimental noise levels. Most significantly, the framework simultaneously infers multiple fundamental processes and properties, including diffusivity, reaction kinetics, chemical potential, and spatial heterogeneity maps, all from mechanical information alone. This method establishes a paradigm for materials characterization, enabling accurate learning of chemical processes in energy storage systems, catalysts, and phase-change materials where conventional diagnostics prove difficult. By revealing that mechanical deformation patterns serve as information-rich `fingerprints' of the underlying chemical processes, this work follows the pathway of inversely learning constitutive laws, with broad implications in materials science and engineering. 
\end{abstract}

\keywords{Chemomechanics \and PDE-constrained optimization \and Mass transport \and Phase separation \and Heterogeneous reaction rate}

\section*{Introduction}

Across scientific and engineering disciplines, inferring physical properties that are difficult to measure directly from other related observed variables underpins many measurement technologies across scales and domains and remains a persistent challenge. In materials science and mechanics, solving inverse problems enables the characterization of properties that are challenging to access directly. Examples include stress fields reconstructed from surface strain measurements \cite{Rethore2018,cameron2021full} and learning reaction-transport coupling from observed thermal wave dynamics \cite{kim2024learning}. Biological systems similarly exploit this paradigm through sophisticated sensing mechanisms. Tendon mechanics are characterized from tissue deformation measurements \cite{huff2024deep}, and cellular traction forces are inferred from fluorescent protein tracking and substrate displacement analysis \cite{Schmitt2024}. These approaches rely on causal physical relationships between observable signals and hidden properties of interest, establishing measurable proxies for otherwise inaccessible quantities. The common challenge shared between these approaches is learning causal physics from sparse, noisy, and often spatially or temporally limited measurements, typically through physics-based model inversion.

A particularly demanding instance of such an approach arises in coupled chemomechanical systems, where chemical processes induce measurable mechanical responses through chemomechanical coupling. In these systems, mechanical deformations encode implicit information about the underlying chemical dynamics through constitutive relationships that link concentration and strain fields. For example, intercalation-induced stresses in lithium-ion battery electrodes directly reflect spatially varying lithium concentration fields and reaction kinetics \cite{Song2025,xu2021guiding}. Hydrogen embrittlement in structural alloys manifests as localized strain concentrations that trace hydrogen diffusion pathways \cite{Robertson2015,ulmer1991hydrogen}, while concrete degradation due to alkali–silica reactions produces expansion strains that encode the kinetics of gel formation and transport processes \cite{rajabipour2015alkali}. In soft matter, chemomechanical coupling governs nuclear condensate morphology and chromatin organization \cite{Zhao2025Condensate}, drives large deformations and pattern dynamics in biological shells \cite{Yin2022Chiral}, and regulates the mechanics of proteins and DNA(deoxyribonucleic acid) in living cells \cite{bao2003cell, Strom2024Condensate}.

Advanced characterization techniques, including in situ X-ray diffraction, neutron scattering, and nuclear magnetic resonance spectroscopy, provide direct chemical measurements and have advanced our understanding of material systems \cite{deng2022correlative, grey2004nmr, Park2021Fictitious}. However, many applications would benefit from complementary approaches that exploit mechanical strain field or displacement measurements to characterize chemical processes that elude direct observation. Such mechanical measurements can be readily obtained across multiple length scales: at the nanoscale through scanning transmission electron microscopy \cite{savitzky2021py4dstem} and Bragg coherent diffractive imaging \cite{sun2024operando}, and at the macroscale via digital image correlation (DIC) \cite{Sutton2009, qi2010in} and fluorescent DIC \cite{jones2014situ,xie2021insitu}.  Additionally, methods that integrate seamlessly with existing characterization workflows and can be applied to complex geometries or operating environments where traditional techniques may be challenging to implement are of interest. Recent advances in computational inverse methods have shown promise for learning constitutive laws from data. Partial differential equation (PDE)-constrained optimization has been applied to infer material laws \cite{Zhao2020, Zhao2021, Zhao2023, deng2022correlative, akerson2025learning, kim2024learning}, and physics-informed neural networks enable solutions to inverse problems in multiphysics systems \cite{Raissi2019,karniadakis2021physics}. Image-based inversion further demonstrates that spatiotemporal thermography and digital image correlation can extract thermomechanical properties, such as thermal diffusivity and expansion coefficients, in a non-destructive, data-driven manner \cite{cottrill2020simultaneous}. Nonetheless, the challenge of inferring chemical kinetics from mechanical signatures in coupled systems remains largely unexplored, highlighting \textit{inferring chemistry from mechanics} as a promising complementary approach for material characterization.

\begin{figure}[htbp]  
\centering
\includegraphics[width=1\linewidth]{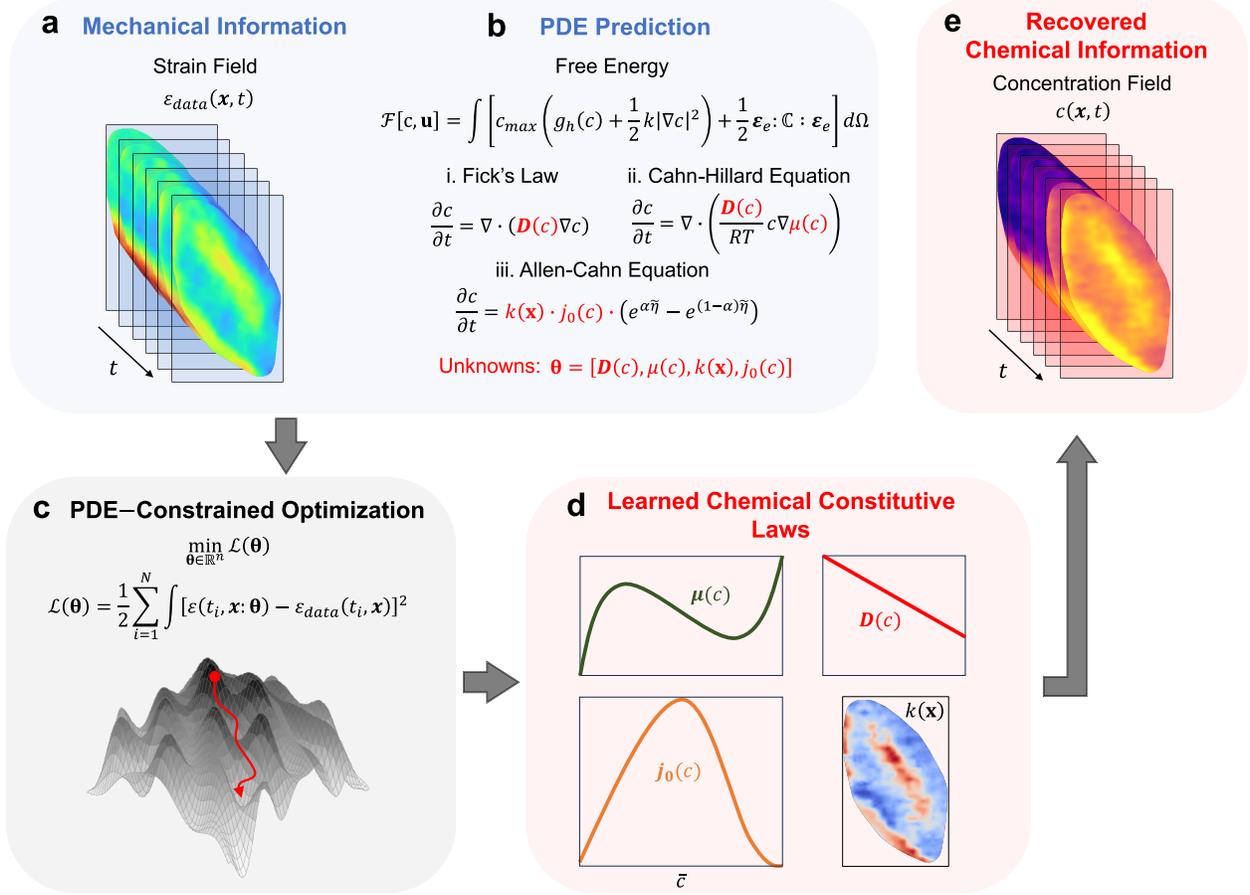}
\caption{
    \textbf{Partial differential equation (PDE)-constrained optimization framework for learning reaction-diffusion kinetics from mechanical information.} 
    \textbf{a} Spatiotemporal strain fields obtained via full-field imaging of battery electrode nanoparticles undergoing charge/discharge half cycles. 
    \textbf{b} Governing equations for the forward model operator with three distinct laws governing species transport, with unknown properties to be learned. 
    \textbf{c} Partial differential equation (PDE)-constrained optimization that minimizes the squared sum of errors at all data points subject to model constraints, with respect to the unknowns. 
    \textbf{d} Learned unknown chemical constitutive laws, such as the concentration-dependent diffusivity, chemical potential,  and exchange current, as well as the heterogeneous reaction rate. 
    \textbf{e} Recovered spatiotemporal mapping of lithium concentration in the nanoparticle.
}
\label{fig:Figure1}
\end{figure}

In this study, we show that mechanical information from spatiotemporal strain measurements can serve as a proxy for learning chemical constitutive laws. Building on advances in high-resolution in situ imaging and physics-informed computational methods, we present a framework for inferring diffusion–reaction kinetics directly from strain fields. Our approach combines PDE-constrained optimization with image-based inversion, enabling physically interpretable recovery of chemical constitutive laws (Fig.~\ref{fig:Figure1}). We implement and validate this methodology in battery electrode materials, where reaction–diffusion-induced strain fields reveal underlying lithium transport and intercalation kinetics. We examine three representative cases: classical Fickian diffusion, spinodal decomposition with pattern formation, and heterogeneous electrochemical reactions with spatially varying rates. Battery electrodes provide an ideal model system due to their well-established chemomechanical coupling, practical relevance for energy storage, and availability of complementary characterization techniques for validation.

\section*{Results}

\subsection*{Mechanical information encodes hidden chemical transport law}
\label{subsec:Ficks}

\begin{figure}[htbp]  
\centering
\includegraphics[width=1\linewidth]{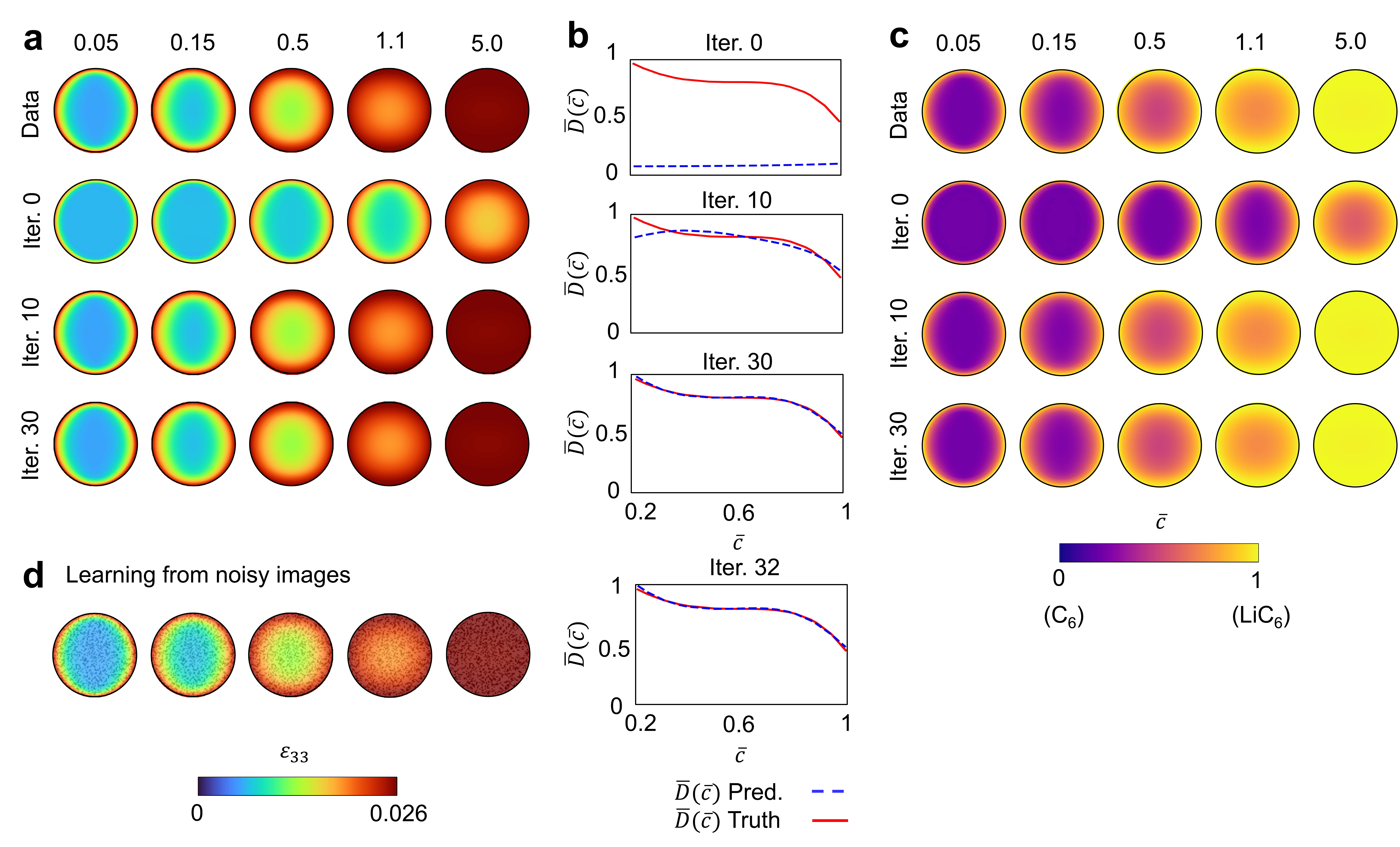}
\caption{
    \textbf{Learning Fickian transport law from spatiotemporal strain fields.}  
    \textbf{a} Surface strain field evolution during the inversion of concentration-dependent diffusivity $\bar{D}(\bar{c})$. The top row shows five training snapshots and subsequent rows show reconstructed fields at key optimization iterations, starting from the initial guess at iteration 0. Numbers above the frames indicate elapsed time (seconds) since the initial frame. 
   \textbf{b} Convergence of the inferred diffusivity $\bar{D}(\bar{c})$ across iterations.
    \textbf{c} Surface concentration evolution during the inversion of concentration-dependent diffusivity $\bar{D}(\bar{c})$. The top row shows ground-truth fields at five time points (not used in training). Subsequent rows show reconstructed fields progressing toward convergence with ground truth data. 
    \textbf{d} Robustness assessment of learning $\bar{D}(\bar{c})$ using training images corrupted with Gaussian noise of standard deviation, $\sigma_N=3 \times 10^{-4}$.
}
\label{fig:Figure2}
\end{figure}

We first demonstrate our framework using a graphite electrode system where Li-ion transport follows Fickian diffusion. The framework was trained on five temporal snapshots of full-field strain tensor components ($\boldsymbol{\varepsilon}_{11}$, $\boldsymbol{\varepsilon}_{22}$, $\boldsymbol{\varepsilon}_{12}$, and $\boldsymbol{\varepsilon}_{33}$) obtained from simulated half-cycle Li-ion intercalation with known ground truth parameters (see Supplementary Note 1 for model parametrization). The surface strain evolution is shown in Fig.~\ref{fig:Figure2}a. Convergence was reached quickly, with the concentration-dependent diffusivity $D(\bar{c})$ (where $\bar{c}=c/c_{\text{max}}$) substantially refined within 10 iterations and fully converged by 30 iterations. The learned diffusivity (Fig.~\ref{fig:Figure2}b), concentration fields (Fig.~\ref{fig:Figure2}c), and their radial profiles (Supplementary Fig.~1) match the ground truth at convergence, verifying the physical consistency of our approach.

Furthermore, real-world implementation demands robustness to the available spatial information and noise in experimental measurements. We evaluated the framework's performance under conditions that mirror practical imaging scenarios. When the training data were restricted to a small subregion containing only 200 nodal points, merely 5\% of the entire domain, the learned diffusivity profile remained in excellent agreement with the ground truth (Supplementary Fig.~2). This demonstrates that localized mechanical information can effectively constrain global chemical transport laws, enabling experimental applications where measurements on the entire domain are impractical. To analyze the influence of experimental noise, we consider realistic measurement conditions based on reported strain measurement capabilities. Considering practical limitations including finite signal-to-noise ratios, material heterogeneity, and measurement uncertainties, we adopt three noise scenarios, a baseline noise standard deviation of $\sigma_N=3 \times 10^{-4}$ representing realistic experimental conditions based on the demonstrated detection of sub-percent strain changes \cite{ulvestad2015topological, savitzky2021py4dstem}, and two over-noise cases with $\sigma_N=1 \times 10^{-3}$ and $\sigma_N=5 \times 10^{-3}$ to test the robustness of the framework under increasingly challenging measurement conditions. The framework demonstrates robustness in reconstructing the diffusivity $D(\bar{c})$ across the noise conditions (Fig.~\ref{fig:Figure2}d and Supplementary Fig.~3). Although fidelity is exhibited, a modest reduction in inversion accuracy becomes discernible at the most severe noise condition of $\sigma_N=5 \times 10^{-3}$ (Supplementary Fig.~3).

\begin{figure}[htbp]  
\centering
\includegraphics[width=1\linewidth]{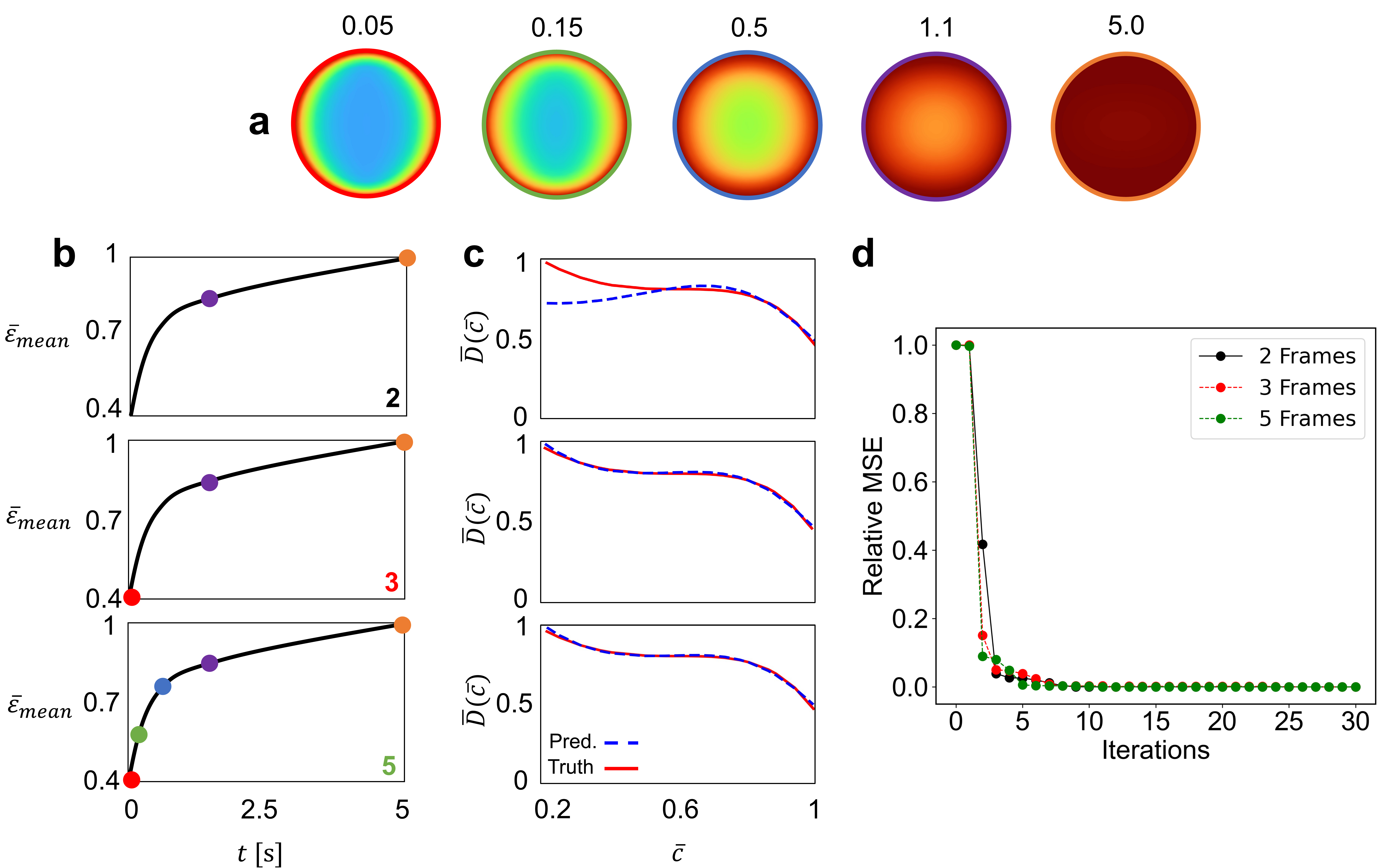}
\caption{
    \textbf{Effect of temporal information loss on learning Fickian transport law from spatiotemporal strain fields.}  
    \textbf{a} Candidate surface strain field snapshots used for case-wise inversion of concentration-dependent diffusivity $\bar{D}(\bar{c})$.Numbers above the frames indicate elapsed time (seconds) since the initial frame. \textbf{b} Corresponding average strain vs time ($\bar{\boldsymbol{\varepsilon}}_{mean}$ -- $\bar{t}$) trajectories showing the temporal locations of selected training images. \textbf{c} Learned diffusivity $\bar{D}(\bar{c})$ profiles at convergence for each test case. \textbf{d} Evolution of the relative mean squared error (MSE) ($\mathcal{L}_i/\mathcal{L}_\text{max}$) during training for each case.
}
\label{fig:Figure3}
\end{figure}

We also examined two factors critical for accurate inference: temporal information content and the fidelity of the mechanical constitutive law. The results show that the initial and final time frames provide the strongest constraints on diffusivity reconstruction by bounding the spatiotemporal evolution of the concentration field though the strain field (\ref{fig:Figure3}). These temporal boundaries capture the system at its most distinct states: the initial conditions establish the initial concentration distribution and driving gradients, while the final state represents the equilibrium or quasi-equilibrium configuration that the diffusion process approaches. The intermediate time steps, while providing additional information about the transient dynamics, contribute less to constraining the inverse problem due to the smoothing nature of diffusive processes. This suggests that strategic temporal sampling can optimize information extraction from limited experimental datasets. Furthermore, perturbing the transverse ($C_{11}$) and axial ($C_{33}$) stiffness components by 10\% induced substantial deviations in the inferred diffusivity profile (Supplementary Fig.~4), highlighting sensitivity to mechanical constitutive assumptions. This chemomechanical coupling presents both challenges and opportunities for materials characterization. To address this limitation, we carried out a joint inference approach that simultaneously learns diffusivity $D(\bar{c})$ and critical stiffness components. When restricted to the principal elastic moduli governing deformation along the crystallographic $a$- and $c$-axes (Supplementary Fig.~5a) ($C_{11}$ and $C_{33}$), the inversion achieved high fidelity (Supplementary Fig.~6), consistent with anisotropic mechanical response during lithium intercalation. However, including additional stiffness components resulted in reduced inversion accuracy (Supplementary Fig.7). While recent studies show that anisotropic properties can be recovered from full-field strain data in purely mechanical systems~\cite{ihuaenyi2024seeking,ihuaenyi2025mechanics}, these results reveal fundamental identifiability challenges in coupled multiphysics systems.

\subsection*{Learning pattern formation from mechanical information}
\label{subsec:CH}

Phase separation in solids represents one of nature's most ubiquitous pattern-forming processes, widely observed in physical, chemical, and biological systems. Yet the underlying chemical driving forces, particularly the interplay between diffusion kinetics and thermodynamic instabilities, remain challenging to measure directly during the rapid evolution of spatially complex patterns. We demonstrate that our framework can decode these hidden dynamics by learning the constitutive laws governing spinodal decomposition from mechanical signatures. Our approach targets systems in which phase separation is driven by chemical potential gradients, with concentration field evolution governed by the Cahn-Hilliard equation \cite{Cahn1958,Cahn1965}.

\begin{figure}[htbp]  
\centering  
\includegraphics[width=1\linewidth]{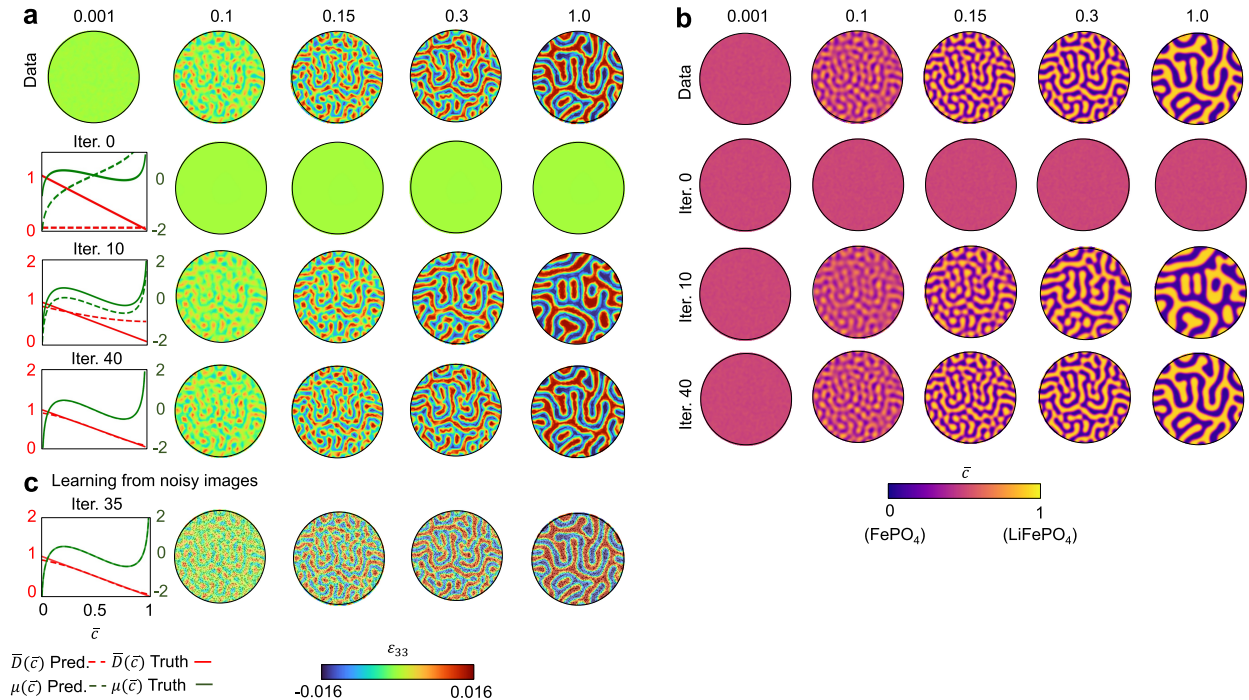}  
\caption{  
    \textbf{Learning pattern formation from spatiotemporal strain fields.}  
    \textbf{a} Evolution of surface strain fields and convergence of the inferred diffusivity $D(\bar{c})$ and chemical potential $\mu(\bar{c})$ during phase separation. The top row displays five training images sampled at distinct time points, while subsequent rows show the reconstructed strain fields at key optimization iterations, beginning from a non-pattern-forming initial guess (iteration 0). Numbers above the frames indicate elapsed time (seconds) since the initial frame.
    \textbf{b} Reconstruction of the underlying concentration fields at key stages of the inversion, demonstrating convergence toward the ground truth distribution by iteration 40.
    \textbf{c} Inference of $D(\bar{c})$ and $\mu(\bar{c})$ from training images corrupted with Gaussian noise of standard deviation $\sigma_N=3 \times 10^{-4}$.
} 
\label{fig:Figure4}  
\end{figure}

Trained on five temporal snapshots of strain fields during pattern formation, the inverse learning framework converged from a completely non-patterned initial guess to the ground truth constitutive laws within 40 iterations. At convergence, the diffusivity $D(\bar{c})$ and the chemical potential $\mu(\bar{c})$ are accurately learned. At the final iteration, the predicted strain fields (Fig.~\ref{fig:Figure4}a) and the reconstructed concentration fields (Fig.~\ref{fig:Figure4}b) showed excellent agreement with the ground truth data. Also, Supplementary Figure~8 presents quantitative comparisons between additional strain field components used in training $\varepsilon_{11}$, $\varepsilon_{22}$, and $\varepsilon_{12}$ and their predictions, demonstrating excellent agreement. This result demonstrates that the strain field contains sufficient information to characterize the thermodynamic and kinetic drivers of phase separation.

\begin{figure}[htbp]  
\centering  
\includegraphics[width=0.85\linewidth]{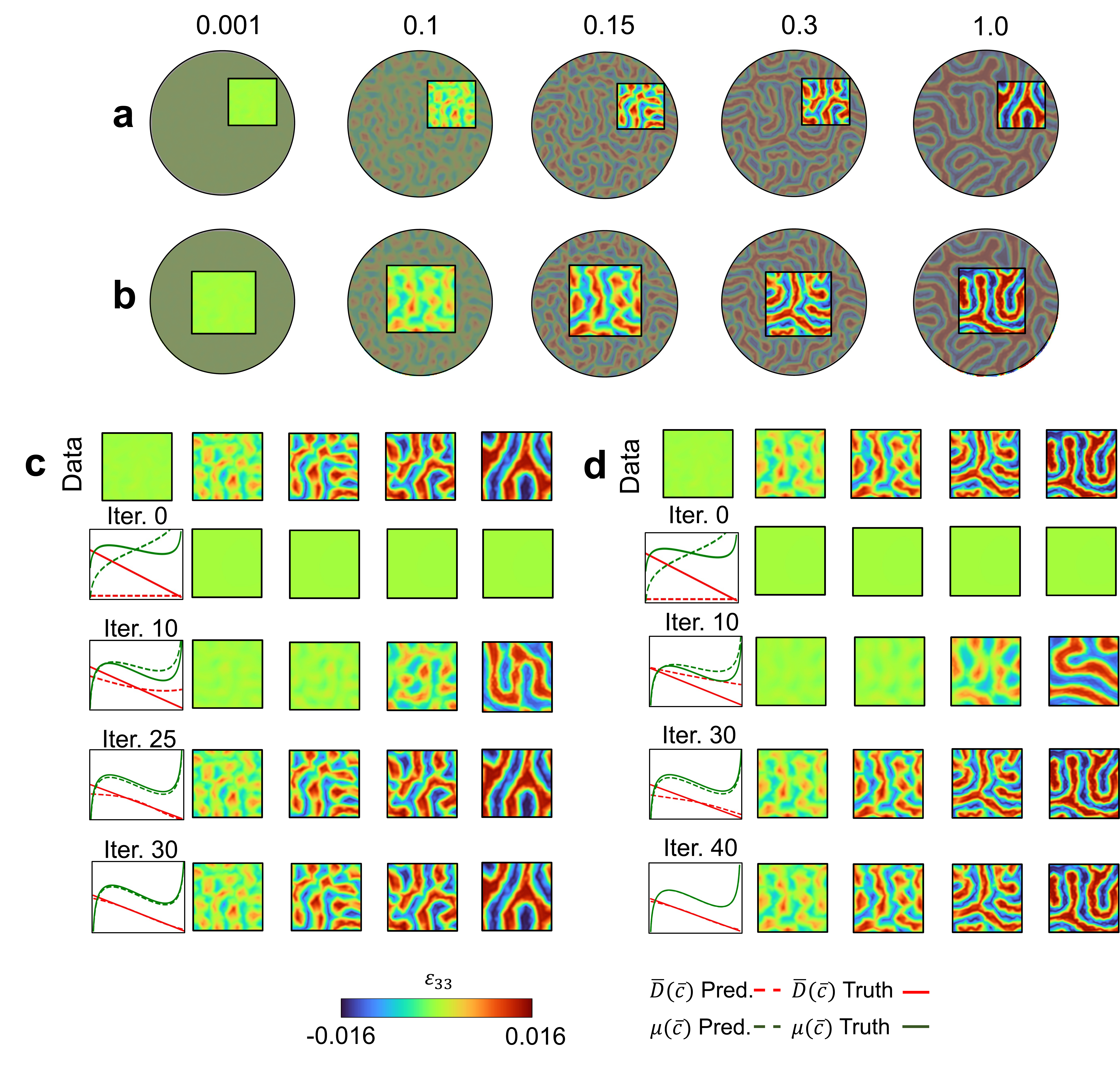}  
\caption{  
    \textbf{Effect of spatial information loss on learning pattern formation from spatiotemporal strain fields.}  
    \textbf{a} Strain field snapshots from the corner of the domain, extracted from a larger simulation domain and used for inversion.Numbers above the frames indicate elapsed time (seconds) since the initial frame.
\textbf{b} Strain field snapshots from the center of the domain, similarly extracted for inversion.
\textbf{c} Evolution of the learned diffusivity $\bar{D}(\bar{c})$ and chemical potential $\mu(\bar{c})$ profiles, shown alongside the corresponding subdomain strain fields extracted from the corner of the domain. The learning process begins with a non-phase-separating initial guess (Iteration 0) and converges by the $30^{th}$ iteration.
\textbf{d} Evolution of the learned diffusivity $\bar{D}(\bar{c})$ and chemical potential $\mu(\bar{c})$ profiles, shown alongside the corresponding subdomain strain fields extracted from the center of the domain. The learning process begins with a non-phase-separating initial guess (Iteration 0) and converges by the $40^{th}$ iteration.
} 
\label{fig:Figure5}  
\end{figure}

Furthermore, the complexity of phase separation patterns demands high measurement fidelity and spatiotemporal sampling. The framework demonstrated robustness under realistic and elevated experimental noise conditions, accurately learning both diffusivity $D(\bar{c})$ and chemical potential $\mu(\bar{c})$ (Fig.~\ref{fig:Figure4}c, Supplementary Fig.~9). However, the inversion of $\mu(\bar{c})$ proved more accurate than $D(\bar{c})$, with the inversion accuracy of the diffusivity degrading progressively with increasing noise amplitude (Supplementary Fig.~9). Spatial information requirements revealed surprising flexibility. Training on spatially restricted subdomains enabled accurate inversion of the governing physics of phase separation, as shown in Figure~\ref{fig:Figure5}. During the inversion process, strain field snapshots were extracted from subdomains within the larger simulation domain: from a corner region (Fig.~\ref{fig:Figure5}a) and from the center of the domain (Fig.~\ref{fig:Figure5}b). The inversion results for both test cases accurately inferred the unknown diffusivity $\bar{D}(\bar{c})$ and chemical potential $\mu(\bar{c})$ profiles (Figs.~\ref{fig:Figure5}a and \ref{fig:Figure5}b). This result demonstrates that localized strain measurements can constrain global physical laws through the encoded pattern formation signatures.

Temporal sampling proved critical for accurate inversion. Different phases of spinodal decomposition encode distinct physical information essential for learning constitutive laws. Five temporal snapshots improved accuracy (Supplementary Fig.~10), with each contributing unique constraints: early times capture thermodynamic instability onset, intermediate states reveal nonlinear fluctuation amplification and pattern connectivity, and late times reflect coarsening dynamics toward equilibrium. While technically feasible with two snapshots, minimal sampling caused substantial inversion inaccuracies, particularly for the chemical potential $\mu(\bar{c})$. This sensitivity reflects the role of the chemical potential in encoding the energy landscape driving phase separation. Capturing this information accurately requires observing the complete nucleation-to-equilibrium evolution. Effective experimental application requires dense temporal sampling throughout this transition.

\subsection*{Learning heterogeneous reaction kinetics in phase separating solids}
\label{subsec:AC}

Battery electrodes and fuel cell catalysts often exhibit spatially heterogeneous reaction rates, particularly in unstable interphases, where direct quantification is challenging \cite{Zhao2023, Lim2016, Mefford2021}. The coupled effects of reaction kinetics, surface chemistry, and phase separation complicate inference. Here, we show that our framework can simultaneously uncover heterogeneous kinetics and phase transformation dynamics from measured strain fields. This capability addresses a fundamental challenge of understanding how spatial heterogeneities in reaction rates couple with thermodynamic instabilities to control pattern formation and performance in functional materials. 

In this case study, we applied the framework to learn heterogeneous reaction kinetics from scanning transmission X-ray microscopy (STXM) data of a carbon-coated lithium iron phosphate (LFP) nanoparticle. We utilized data from STXM imaging of an LFP platelet specimen positioned within a microfluidic electrochemical setup \cite{deng2022correlative, Zhao2023}. The electrochemical process involves lithium ion intercalation into the LFP structure during discharge (Li$^+$ + FePO$_4$ + e$^-$ $\rightarrow$ LiFePO$_4$). The STXM measurements provide spatially-resolved X-ray absorption data, which were converted to lithium concentration distributions averaged through the [010] crystallographic direction (b-direction). Our analysis focuses on experimental data acquired during a single discharge half-cycle at 0.6C rate, comprising a temporal series of six STXM images that capture lithium insertion dynamics. The strain field evolution during discharge was reconstructed using a thermodynamically consistent chemomechanical framework \cite{Zhao2023}.

Additionally, the particles exhibit spatially varying local reaction rates caused by microstructural inhomogeneities. The temporal evolution of concentration distributions follows a depth-averaged, reaction-limited Allen-Cahn framework \cite{Bazant2013, Bazant2017}, which describes the fundamental physics governing the competition between thermodynamic driving forces and reaction kinetics. To solve the inverse problem, we simultaneously learn the concentration-dependent exchange current $j_0(c)$ that characterizes the intrinsic reaction kinetics, the chemical potential $\mu(c)$, and the spatial heterogeneity prefactor $k(\mathbf{x})$ that controls local reaction rate variations.

\begin{figure}[htbp]  
\centering  
\includegraphics[width=1\linewidth]{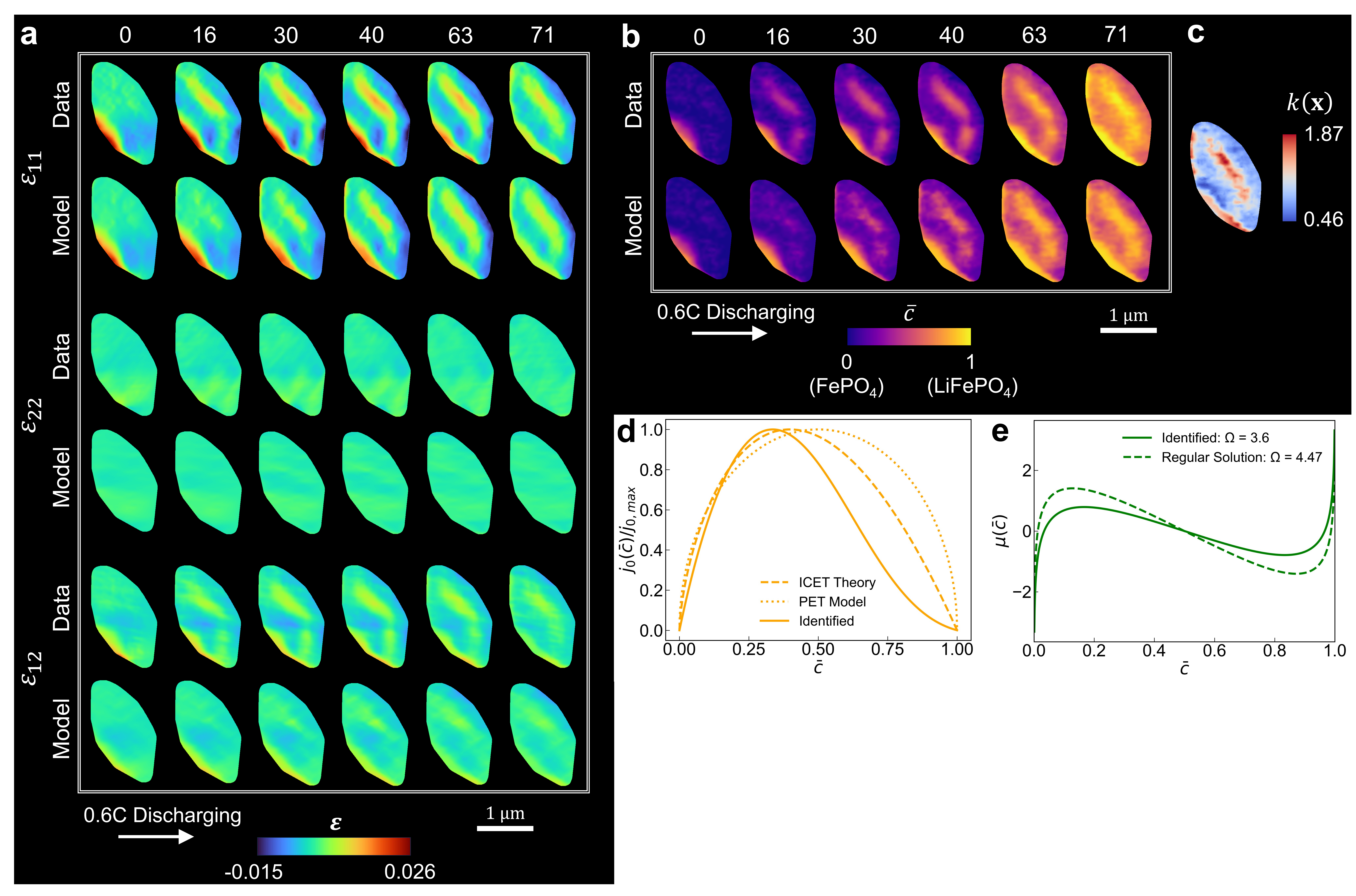}  
\caption{
\textbf{Learning heterogeneous reaction kinetics from spatiotemporal strain fields.}  
\textbf{a} Experimental and simulated strain fields at key time frames. Rows labeled `Data' and `Model' represent experimental measurements and model predictions using the learned constitutive laws, respectively. Numbers above the frames indicate elapsed time (minutes) since the initial half-cycle frame. Predicted strain fields demonstrate close agreement with experimental observations.  
\textbf{b} Experimental and simulated lithium-ion concentration fields at corresponding time points. The reconstructed concentration fields agree well with the experimental data, validating the framework. 
\textbf{c} The learned spatial heterogeneity map $k(\mathbf{x})$ across the particle surface, accounting for variations in the local reaction rates. 
\textbf{d} Comparison of the learned exchange current with theoretical predictions. The solid line shows the inversion result for normalized exchange current $j_0(\bar{c})/j_{0,\mathrm{max}}$. Results are benchmarked against ICET theory $j_0(\bar{c})=\bar{c}^{0.66}(1-\bar{c})$~\cite{Fraggedakis2021} (dashed line), which extends classical
charge transfer formalism to incorporate ion transfer limitations, and the symmetric form $j_0(\bar{c})=\sqrt{\bar{c}(1-\bar{c})}$ from classical porous electrode theory~\cite{doyle1993modeling} (dotted line). 
\textbf{e} Learned chemical potential (solid line) compared with the regular solution model using $\Omega=4.47$ (dashed line).
}
\label{fig:Figure6}  
\end{figure}

Our framework leverages six temporal snapshots of the strain field data ($\varepsilon_{11}$, $\varepsilon_{22}$, and $\varepsilon_{12}$). Quantitative validation reveals close agreement between experimental strain measurements and model predictions using the learned constitutive laws (Fig.\ref{fig:Figure6}a). The reconstructed concentration fields agree well with experimental observations, validating the framework (Fig.\ref{fig:Figure6}b). Also revealed is the spatial heterogeneity map across the particle domain (Fig.\ref{fig:Figure6}c). 

To benchmark our approach against established theory, we compared the derived normalized exchange current with predictions from intercalation-limited electron transfer (ICET) theory, $j_0(\bar{c})=\bar{c}^{0.66}(1-\bar{c})$~\cite{Fraggedakis2021}, which extends classical charge transfer formalism to incorporate ion transfer limitations (Fig. \ref{fig:Figure6}d). This comparison demonstrates a closer agreement with ICET predictions relative to the symmetric exchange current model $j_0(\bar{c})=\sqrt{\bar{c}(1-\bar{c})}$ conventionally employed in Butler-Volmer kinetics within porous electrode theory~\cite{doyle1993modeling}. The asymmetric nature of our inferred exchange current aligns closer with ICET theoretical expectations, whereas classical porous electrode formulations fail to capture this fundamental asymmetry inherent to intercalation processes. The reconstructed chemical potential $\mu(\bar{c})$ (Fig. \ref{fig:Figure6}e) demonstrates qualitative consistency with the established regular solution model for LiFePO$_4$~\cite{cogswell2012coherency}, $\mu_h(\bar{c}) = \ln \left( \bar{c}/\left(1 - \bar{c} \right) \right) + \Omega (1 - 2\bar{c})$ where $\Omega=4.47$, albeit with a reduced nucleation barrier. The inferred $\mu_h(\bar{c})$ exhibits sensitivity to model selection and the gradient energy coefficient $k$, which was fixed to its literature value due to its non-identifiability \cite{Zhao2023}. 

\section*{Discussion}

By demonstrating that mechanical signatures can serve as trustworthy proxies for chemical processes, this work opens new diagnostic avenues where traditional chemical measurements prove difficult. The fundamental insight that mechanical deformation fields encode sufficient information to reconstruct complex reaction-diffusion kinetics reveals a connection between mechanical and chemical domains, where mechanical responses contain rich information about underlying thermodynamic and kinetic landscapes. The progression from Fickian diffusion to phase separation dynamics and heterogeneous reaction kinetics demonstrates framework versatility across increasing complexity levels. Particularly significant is the simultaneous inference of multiple coupled parameters: diffusivity, chemical potential, exchange current, and spatial heterogeneity, from mechanical measurements alone, addressing the challenge of requiring multiple, often incompatible techniques for complete system characterization. For battery systems, demonstrations with graphite and LFP electrode materials establish fundamental physics, informing future diagnostic approaches. The scale-dependent measurement accessibility for mechanical and chemical measurements suggests particular value for composite electrode-level applications where bulk deformation can be tracked non-invasively while internal chemical states remain inaccessible, complementing existing nanoscale techniques.

Although the demonstrated cases highlight commonly used battery electrode materials \cite{Song2025, Ihuaenyi2025, Zhang2021}, the framework's physics-based foundation suggests broad applicability to other chemomechanically coupled systems, including catalytic materials \cite{MartinezAlonso2022}, and chemically responsive polymers and biomaterials \cite{hua2021swaying,ihuaenyi2021orthotropic,girard2007cellular}. This work represents progress toward integrated material characterization, recognizing different physical domains as interconnected information sources rather than isolated targets. However, current limitations must be acknowledged. The framework assumes physics-based models fully describe multiphysics phenomena without information loss. Additionally, in practice, as shown by the results, measurement noise, spatial averaging, and temporal limitations can restrict the fidelity of the inversion. Hence, inference in such coupled systems should also account for uncertainties. Furthermore, the uniqueness and stability of the solution to the inverse problem also depend on the specific coupling between chemical and mechanical processes, which may vary significantly across different material systems. The computational requirements for real-time implementation, particularly for three-dimensional systems with complex geometries, represent another practical consideration. However, advances in computational methods suggest that these limitations are surmountable rather than fundamental.

\section*{Methods}

\subsection*{Governing equations}

We consider a domain \( \Omega \subset \mathbb{R}^2 \) characterized by the species concentration \( c(\mathbf{x}, t) \) and the displacement field \( \boldsymbol{u}(\mathbf{x}, t) \). To model the coupled phenomena of species transport and mechanical deformation, the free energy is expressed as a variational functional consisting of chemical free energy and mechanical energy:
\begin{equation}
\label{eq:Free_E}
\mathcal{F}[c, \mathbf{u}] = \int_{\Omega} \left[ c_{\text{max}}\left( g_h(\bar{c}) + \frac{1}{2} k |\nabla \bar{c}|^2 \right) + \frac{1}{2}\boldsymbol{\varepsilon}_{\text{e}} : \mathbb{C} : \boldsymbol{\varepsilon}_{\text{e}} \right] d\Omega,
\end{equation}

where  $c_{\text{max}}$ is the maximum lithium concentration, $\bar{c}=c/c_{max}$ is the normalized lithium fraction ($0\leq \bar{c} \leq1$), \( g_h(\bar{c}) \) represents the homogeneous chemical free energy density, \( k \) is the gradient energy coefficient accounting for interfacial effects, and \( \nabla \bar{c} \) denotes the concentration gradient. The final term describes the elastic energy stored in the system, where \( \boldsymbol{\varepsilon}_e \) is the elastic strain tensor, \( \mathbb{C} \) is the fourth-rank elasticity tensor, and \( \boldsymbol{\varepsilon}_e: \mathbb{C}: \boldsymbol{\varepsilon}_e \) represents the double contraction of these tensors.

For a deformable body that occupies the domain \( \Omega \), the strain-displacement relationship considering the assumption of small deformations is given by:
\begin{equation}
\label{eq:kinematics}
\boldsymbol{\varepsilon}(\mathbf{u}) = \frac{1}{2} \left( \nabla \mathbf{u} + \nabla \mathbf{u}^\intercal \right), \quad \text{in } \Omega.
\end{equation}

The total strain tensor \( \boldsymbol{\varepsilon} \) is additively decomposed into mechanical and chemical contributions:
\begin{equation}
\label{eq:strain_decomp}
\boldsymbol{\varepsilon} = \boldsymbol{\varepsilon}_{mech} + \boldsymbol{\varepsilon}_{ch},
\end{equation}

where the mechanical strain \( \boldsymbol{\varepsilon}_{mech} \) is purely elastic under small deformation assumption, i.e., \( \boldsymbol{\varepsilon}_{mech} = \boldsymbol{\varepsilon}_{e} \). The chemical strain \( \boldsymbol{\varepsilon}_{ch} \) arises from compositional changes and is modeled as an anisotropic volumetric expansion proportional to the local concentration variation:
\begin{equation}
\label{eq:chem_strain}
\boldsymbol{\varepsilon}_{ch} = \boldsymbol{\beta} (c - c_0),
\end{equation}

where \( \boldsymbol{\beta} \) is the chemical expansion tensor, \( c \) is the local concentration and \( c_0 \) is the reference concentration.

Assuming quasistatic mechanical equilibrium, the variational condition for mechanical equilibrium is:
\begin{equation}
\label{eq:variation_u}
\frac{\delta \mathcal{F}}{\delta \mathbf{u}} = \nabla \cdot \boldsymbol{\sigma} = 0.
\end{equation}

Then, the Cauchy stress tensor \( \boldsymbol{\sigma} \) is derived from the constitutive relation:
\begin{equation}
\label{eq:constitutive}
\boldsymbol{\sigma} = \mathbb{C} : (\boldsymbol{\varepsilon} - \boldsymbol{\varepsilon}_{ch}).
\end{equation}

Furthermore, in the limit of chemical equilibrium, the chemical potential \( \mu \) is defined as the variational derivative of the free energy:
\begin{equation}
\label{eq:chem_potential}
\mu = \frac{\delta \mathcal{F}}{\delta c} = \mu_h(\bar{c}) - k \nabla^2 \bar{c} - \boldsymbol{\beta}:\boldsymbol{\sigma},
\end{equation}

where \( \mu_h(\bar{c}) \) is the homogeneous part of the chemical potential, \( -k \nabla^2 \bar{c} \) represents the gradient energy contribution.

The transport of species is governed by a diffusion process, as described by the conservation of species:
\begin{equation}
\label{eq:continuity}
\frac{\partial c}{\partial t} + \nabla \cdot \mathbf{J} = 0,
\end{equation}

where \( \mathbf{J} \) is the species flux. Based on kinetic theory, the diffusion flux is proportional to the gradient of the chemical potential:
\begin{equation}
\label{eq:flux}
\mathbf{J} = -\mathbf{M} c \nabla \mu,
\end{equation}

where \( \mathbf{M} = \mathbf{D}(c) \left(RT\right)^{-1} \) is the orientation-dependent mobility tensor, \( R \) is the universal gas constant, \( T \) is the absolute temperature, and \( \mathbf{D}(c) \) is the orientation and concentration-dependent diffusivity tensor. This formulation gives us the Cahn-Hilliard equation, which describes phase separation driven by chemical potential gradients \cite{Cahn1958, Cahn1965}. For ideal diffusion, where the chemical potential is linearly related to the concentration, the flux is simplified as:
\begin{equation}
\label{eq:flux_diffusion}
\mathbf{J} = -\mathbf{D}(c) \nabla c.
\end{equation}

Additionally, to model the evolution of lithium concentration within thin platelet particles such as LFP, a reaction-limited Allen-Cahn reaction model is typically used\cite{Bazant2013,Bazant2017}. The Allen-Cahn model is expressed as:
\begin{equation}
\label{eq:allen_cahn}
\frac{\partial c}{\partial t} = k(\mathbf{x})R(c,\eta),
\end{equation}

where $k(\mathbf{x})$ for $\mathbf{x} = (x, y) \in \mathbb{R}^2$, is a surface heterogeneity prefactor, $R$ is the reaction rate for a spatially uniform system dependent on the concentration and overpotential, $\eta$. Furthermore, the reaction rate is modeled using Butler-Volmer kinetics:
\begin{equation}
\label{eq:butler_volmer}
R(c,\eta) = j_0(c) \left[ \exp\left( - \alpha \tilde{\eta} \right) - \exp\left( (1-\alpha )\tilde{\eta} \right) \right],
\end{equation}

where $j_0(c)$ is the exchange current normalized by the Faraday constant, $\alpha$ is the symmetry factor ($\alpha=0.5$ \cite{Bai2014}). $\tilde{\eta}$ is the nondimensionalized overpotential defined as:
\begin{equation}
\label{eq:overpotential}
\tilde{\eta} = \frac{\mu-\mu_{res}}{k_{B}T},
\end{equation}

where $\mu_{res}=\mu_{\text{Li}^{+}}-e\nabla\phi$ is the reservoir chemical potential, with $\mu_{\text{Li}^{+}}$ and $\nabla\phi$ representing the chemical potential of lithium ions in the electrolyte and the interfacial voltage, respectively. Also, $k_B$ is the Boltzmann constant and $T$ is the absolute temperature.

\subsection*{Problem setup}

We investigate lithium intercalation during half-cycle reactions across different domain geometries. For Fickian diffusion and Cahn-Hilliard pattern formation, we use a 2D spherical domain ($r=10 \, \mu\text{m}$). For learning heterogeneous reaction kinetics from experimental data, we employ LFP platelet geometry (20$\mu\text{m}$ × 35$\mu\text{m}$) reconstructed from STXM images. Zero-displacement boundary conditions on domain surfaces induce internal mechanical strains during species diffusion. Without mechanical constraints, the system remains stress-free with strain consisting solely of chemical contributions ($\boldsymbol{\varepsilon} = \boldsymbol{\varepsilon}_{ch}$).

For concentration-driven diffusion, chemical strain is proportional to the concentration. Substituting this into the continuity (Eq.~\ref{eq:continuity}) and flux (Eq.~\ref{eq:flux_diffusion}) equations yields the strain evolution equation:
\begin{equation}
    \frac{\partial \boldsymbol{\varepsilon}}{\partial t} + \nabla \cdot \left(-\boldsymbol{D}(c) \nabla \boldsymbol{\varepsilon}\right) = 0.
\label{eq:strain_diffusion_equation}
\end{equation}

Direct diffusivity inference from Eq.~\ref{eq:strain_diffusion_equation} would require an isolated measurement of the chemical strain. This would require an a priori knowledge of the chemical expansion tensor and the spatiotemporal concentration field. However, in real systems, domains are typically subject to mechanical constraints that induce mechanical strains and, consequently, stresses. In such cases, the measured strain can be assumed to be a superposition of mechanical and chemical strains(Eq.~\ref{eq:strain_decomp}). To decouple the measured strain fields and infer diffusion dynamics using only the chemical strain (as formulated in Eq.~\ref{eq:strain_diffusion_equation}), one would need to measure the stress field to isolate the elastic strain. However, measuring stress is infeasible, rendering the problem inherently ill-posed. The governing equations, along with the boundary conditions, define a well-posed boundary value problem (BVP) when the constitutive parameters are known. In this work, we infer reaction-diffusion kinetics directly from spatiotemporal strain fields without a priori knowledge of the concentration field by formulating inverse problems for three different cases. This is achieved by minimizing an objective function defined as the \( L_2 \) norm of the discrepancy between the observed and predicted strain fields:
\begin{equation}
\label{eq:obj_func}
    \mathcal{L}(\boldsymbol{\theta}) = \frac{1}{2} \sum_{i=1}^{N} \int_{\Omega} \left\| \boldsymbol{\varepsilon}(\mathbf{x}, t_i; \boldsymbol{\theta}) - \boldsymbol{\varepsilon}_{\text{data}}(\mathbf{x}, t_i) \right\|^2 \, d\mathbf{x},
\end{equation}

where \( N \) represents the number of training snapshots acquired at discrete time steps \( t_i \), \( \boldsymbol{\varepsilon}_{\text{data}} \) denotes the observed strain field data, and \( \boldsymbol{\varepsilon}(\mathbf{x}, t_i; \boldsymbol{\theta}) \) is the predicted strain field, parameterized by \( \boldsymbol{\theta} \). The minimization of the Eq.\ref{eq:obj_func} is carried out using trust-region constrained optimization (see Supplementary Note 3). The functions \( D(\bar{c}) \), \( \mu_h(\bar{c}) \), and \( j_0(\bar{c}) \) are approximated using a spectral expansion in terms of Legendre polynomials in the forms:
\begin{equation}
    D(\bar{c}) = \sum_{i=1}^{M} a_i P_i(\bar{c}),
\end{equation}
\begin{equation}
    j_0(\bar{c}) = \bar{c}(1-\bar{c})\sum_{i=1}^{M} b_i P_i(\bar{c}),
\end{equation}
\begin{equation}
    \mu_h(\bar{c}) = \log \frac{\bar{c}}{1 - \bar{c}} + \sum_{i=1}^{M} c_i P_i(\bar{c}),
\end{equation}
where \( M\) is the number of terms in the expansion. 
This well-conditioned parametrization leverages the ideal entropy of mixing to enforce physically meaningful bounds on the concentration (\( 0 < \bar{c} < 1 \)). Here, \( P_i(\bar{c}) \) are Legendre polynomials defined on the interval \([0, 1]\), and \( a_i \), \( b_i \), and \( c_i \) are the expansion coefficients to be learned. Legendre polynomials are chosen over standard polynomial bases due to their orthogonality, which mitigates numerical instability in the approximation. While this approach does not guarantee exact recovery of the true forms of \( D(\bar{c}) \), \( \mu_h(\bar{c}) \), and \( j_0(\bar{c}) \), it ensures thermodynamic consistency by enforcing positivity constraints \cite{Callen1985, Bazant2017}. Additionally, the surface heterogeneity prefactor was modeled as a Gaussian random field, parameterized by its eigenvalues $\lambda_i$ and spatial weights $\boldsymbol{\alpha}$ (see see Supplementary Note 2). 

We evaluate our framework through synthetic and experimental validation. Synthetic tests use strain field data generated from governing equations with predetermined parameters (see Supplementary Note 1, Supplementary Figs. S5, S11-S13) for testing the framework in learning Fickian diffusion and pattern formation. In synthetic test cases, we leverage scaling relationships where the characteristic timescale is $D \frac{d\mu}{dc}$ and length scale is $\sqrt{\frac{k}{\mu'_h(c)}}$ to determine constitutive quantities that define the governing PDEs.

For experimental validation, we apply our methodology to STXM imaging data of carbon-coated LiFePO$_4$ nanoparticles during 0.6C discharge. We simultaneously infer chemical potential $\mu_h(\bar{c})$, exchange current density $j_0(\bar{c})$, and spatial heterogeneity $k(\mathbf{x})$ from strain field measurements, validating the learned concentration fields against STXM data.

\section*{Data availability}
Data supporting the verification studies are available within the manuscript and/or the Supporting Information. Experimental STXM  image data used in this study can be found in \url{https://doi.org/10.6084/m9.figshare.30037894.v1}. 

\section*{Code availability}
The simulation and computational codes used in this study are available from the corresponding author upon reasonable request.

\section*{Acknowledgements}

This research was supported by the U.S. National Science Foundation through grant CMMI-2450006.

\section*{Author contributions statement}

R.C.I., H.Z., and J.Z. conceived and designed the study; R.C.I. and R.F. developed computational models; R.C.I. developed software; R.C.I. and H.Z. performed data curation and validation; R.C.I. and J.Z. wrote the manuscript; R.C.I., H.Z., R.F., R.B., M.Z.B., and J.Z.  reviewed and edited the manuscript. 

\section*{Competing Interests} 

J.Z. co-founded BattSafe Solutions LLC to provide safety-focused testing and technical consulting for batteries. M.Z.B. serves as the Chief Scientist and Co-founder of Lithios, Inc., a startup company working on critical mining, and the Chief Scientific Advisor for Saint-Gobain Research North America.

\bibliographystyle{unsrtnat}
\bibliography{references}  

\begin{thebibliography}{72}
\providecommand{\natexlab}[1]{#1}
\providecommand{\url}[1]{\texttt{#1}}
\expandafter\ifx\csname urlstyle\endcsname\relax
  \providecommand{\doi}[1]{doi: #1}\else
  \providecommand{\doi}{doi: \begingroup \urlstyle{rm}\Url}\fi

\bibitem[Réthoré et~al.(2018)Réthoré, Leygue, Coret, Stainier, and Verron]{Rethore2018}
J.~Réthoré, A.~Leygue, M.~Coret, L.~Stainier, and E.~Verron.
\newblock Computational measurements of stress fields from digital images.
\newblock \emph{International Journal for Numerical Methods in Engineering}, 113\penalty0 (12):\penalty0 1810--1826, 2018.
\newblock \doi{https://doi.org/10.1002/nme.5721}.

\bibitem[Cameron and Tasan(2021)]{cameron2021full}
Benjamin~C Cameron and C~Cem Tasan.
\newblock Full-field stress computation from measured deformation fields: A hyperbolic formulation.
\newblock \emph{Journal of the Mechanics and Physics of Solids}, 147:\penalty0 104186, 2021.
\newblock \doi{https://doi.org/10.1016/j.jmps.2020.104186}.

\bibitem[Kim and Deng(2024)]{kim2024learning}
S.~Kim and S.~Deng.
\newblock Learning reaction-transport coupling from thermal waves.
\newblock \emph{Nature Communications}, 15:\penalty0 9930, 2024.
\newblock \doi{10.1038/s41467-024-54177-2}.
\newblock URL \url{https://doi.org/10.1038/s41467-024-54177-2}.

\bibitem[Huff et~al.(2024)Huff, Earl, Goergen, and O'Connell]{huff2024deep}
Reece~D. Huff, Conner~C. Earl, Craig~J. Goergen, and Grace~D. O'Connell.
\newblock Deep learning enables accurate soft tissue tendon deformation estimation in vivo via ultrasound imaging.
\newblock \emph{Scientific Reports}, 14:\penalty0 18401, 2024.
\newblock \doi{https://doi.org/10.1038/s41598-024-68875-w}.

\bibitem[Schmitt et~al.(2024)Schmitt, Colen, Sala, Devany, Seetharaman, Caillier, Gardel, Oakes, and Vitelli]{Schmitt2024}
Matthew~S. Schmitt, Jonathan Colen, Stefano Sala, John Devany, Shailaja Seetharaman, Alexia Caillier, Margaret~L. Gardel, Patrick~W. Oakes, and Vincenzo Vitelli.
\newblock Machine learning interpretable models of cell mechanics from protein images.
\newblock \emph{Cell}, 187\penalty0 (2):\penalty0 481--494, 2024.
\newblock \doi{https://doi.org/10.1016/j.cell.2023.11.041}.

\bibitem[Song et~al.(2025)Song, Ihuaenyi, Lim, Wang, Li, Fang, Ghamsari, Xu, Lee, and Zhu]{Song2025}
Jing Song, Royal~C. Ihuaenyi, Jungsoo Lim, Zhenting Wang, Weiran Li, Ruqing Fang, Amir~K. Ghamsari, Hao Xu, Young~M. Lee, and Juner Zhu.
\newblock A microstructural electrochemo-mechanical model of high-nickel composite electrodes towards digital twins to bridge the particle and electrode-level characterizations.
\newblock \emph{Energy \& Environmental Science}, 18\penalty0 (7):\penalty0 3129--3147, 2025.
\newblock \doi{DOI https://doi.org/10.1039/D4EE04856C}.

\bibitem[Xu et~al.(2021)Xu, Zhu, Finegan, Zhao, Lu, Li, Hoffman, Bertei, Shearing, and Bazant]{xu2021guiding}
Huanyu Xu, Jiawei Zhu, Donal~P Finegan, Hui Zhao, Xuekun Lu, Weibing Li, Noah Hoffman, Antonio Bertei, Paul Shearing, and Martin~Z Bazant.
\newblock Guiding the design of heterogeneous electrode microstructures for li-ion batteries: microscopic imaging, predictive modeling, and machine learning.
\newblock \emph{Advanced Energy Materials}, 11\penalty0 (19):\penalty0 2003908, 2021.
\newblock \doi{https://doi.org/10.1002/aenm.202003908}.

\bibitem[Robertson et~al.(2015)Robertson, Sofronis, Nagao, Martin, Wang, Gross, and Nygren]{Robertson2015}
Ian~M Robertson, Petros Sofronis, Akihide Nagao, Martin~L Martin, Shuai Wang, David~W Gross, and Karl~E Nygren.
\newblock Hydrogen embrittlement understood.
\newblock \emph{Metallurgical and Materials Transactions A}, 46\penalty0 (6):\penalty0 2323--2341, 2015.
\newblock \doi{https://doi.org/10.1007/s11661-015-2836-1}.

\bibitem[Ulmer and Altstetter(1991)]{ulmer1991hydrogen}
D.~G. Ulmer and C.~J. Altstetter.
\newblock Hydrogen-induced strain localization and failure of austenitic stainless steels at high hydrogen concentrations.
\newblock \emph{Acta Metallurgica et Materialia}, 39\penalty0 (6):\penalty0 1237--1248, 1991.
\newblock \doi{https://doi.org/10.1016/0956-7151(91)90211-I}.

\bibitem[Rajabipour et~al.(2015)Rajabipour, Giannini, Dunant, Ideker, and Thomas]{rajabipour2015alkali}
Farshad Rajabipour, Erika Giannini, Cyrille Dunant, Jason~H. Ideker, and Michael~D.A. Thomas.
\newblock Alkali--silica reaction: Current understanding of the reaction mechanisms and the knowledge gaps.
\newblock \emph{Cement and Concrete Research}, 76:\penalty0 130--146, 2015.
\newblock \doi{https://doi.org/10.1016/j.cemconres.2015.05.024}.

\bibitem[Zhao et~al.(2025)Zhao, Strom, Eeftens, Haataja, Ko{\v s}mrlj, and Brangwynne]{Zhao2025Condensate}
Hongbo Zhao, Amy~R. Strom, Jorine~M. Eeftens, Mikko Haataja, Andrej Ko{\v s}mrlj, and Clifford~P. Brangwynne.
\newblock Condensate-driven chromatin organization via elastocapillary interactions.
\newblock \emph{bioRxiv}, June 2025.
\newblock \doi{10.1101/2025.06.12.659369}.
\newblock Preprint.

\bibitem[Yin et~al.(2022)Yin, Li, and Feng]{Yin2022Chiral}
Shiqiang Yin, Bo~Li, and Xi-Qiao Feng.
\newblock Three-dimensional chiral morphodynamics of chemomechanical active shells.
\newblock \emph{Proceedings of the National Academy of Sciences}, 119\penalty0 (49):\penalty0 e2206159119, 2022.
\newblock \doi{10.1073/pnas.2206159119}.

\bibitem[Bao and Suresh(2003)]{bao2003cell}
Gang Bao and Subra Suresh.
\newblock Cell and molecular mechanics of biological materials.
\newblock \emph{Nature Materials}, 2\penalty0 (11):\penalty0 715--725, 2003.
\newblock \doi{https://doi.org/10.1038/nmat1001}.

\bibitem[Strom et~al.(2024)Strom, Kim, Zhao, Chang, Orlovsky, Ko{\v s}mrlj, Storm, and Brangwynne]{Strom2024Condensate}
Amy~R. Strom, Yoonhee Kim, Hongbo Zhao, Yoon~C. Chang, Nikita~D. Orlovsky, Andrej Ko{\v s}mrlj, Cornelis Storm, and Clifford~P. Brangwynne.
\newblock Condensate interfacial forces reposition dna loci and probe chromatin viscoelasticity.
\newblock \emph{Cell}, 187\penalty0 (19):\penalty0 5282--5297, 2024.
\newblock \doi{10.1016/j.cell.2024.07.034}.

\bibitem[Deng et~al.(2022)Deng, Zhao, Jin, Hughes, Savitzky, Ophus, Fraggedakis, Borb{\'e}ly, Yu, Lomeli, and Yan]{deng2022correlative}
H.D. Deng, H.~Zhao, N.~Jin, L.~Hughes, B.H. Savitzky, C.~Ophus, D.~Fraggedakis, A.~Borb{\'e}ly, Y.S. Yu, E.G. Lomeli, and R.~Yan.
\newblock Correlative image learning of chemo-mechanics in phase-transforming solids.
\newblock \emph{Nature Materials}, 21\penalty0 (5):\penalty0 547--554, 2022.
\newblock \doi{https://doi.org/10.1038/s41563-021-01191-0}.

\bibitem[Grey and Dupré(2004)]{grey2004nmr}
Clare~P. Grey and Nicolas Dupré.
\newblock Nmr studies of cathode materials for lithium-ion rechargeable batteries.
\newblock \emph{Chemical Reviews}, 104\penalty0 (10):\penalty0 4493--4512, 2004.
\newblock \doi{https://doi.org/10.1021/cr020734p}.

\bibitem[Park et~al.(2021)Park, Zhao, Kang, Lim, Chen, Yu, Braatz, Shapiro, Hong, Toney, and Bazant]{Park2021Fictitious}
J.~Park, H.~Zhao, S.~D. Kang, K.~Lim, C.~C. Chen, Y.~S. Yu, R.~D. Braatz, D.~A. Shapiro, J.~Hong, M.~F. Toney, and M.~Z. Bazant.
\newblock Fictitious phase separation in li layered oxides driven by electro-autocatalysis.
\newblock \emph{Nature Materials}, 20\penalty0 (7):\penalty0 991--999, 2021.
\newblock \doi{https://doi.org/10.1038/s41563-021-00936-1}.

\bibitem[Savitzky et~al.(2021)Savitzky, Hughes, Ophus, Zeltmann, Minor, et~al.]{savitzky2021py4dstem}
BH~Savitzky, LA~Hughes, C~Ophus, SE~Zeltmann, AM~Minor, et~al.
\newblock py4dstem: a software package for multimodal analysis of four-dimensional scanning transmission electron microscopy datasets.
\newblock \emph{Microscopy and Microanalysis}, 27\penalty0 (4):\penalty0 712--743, 2021.

\bibitem[Sun et~al.(2024)Sun, Hy, Hua, Wingert, Harder, Meng, Shpyrko, and Singer]{sun2024operando}
Yudong Sun, Sunny Hy, Natnael Hua, Jesse Wingert, Ross Harder, Ying~Shirley Meng, Oleg Shpyrko, and Andrej Singer.
\newblock Operando real-space imaging of a structural phase transformation in the high-voltage electrode li$_x$ni$_{0.5}$mn$_{1.5}$o$_4$.
\newblock \emph{Nature Communications}, 15\penalty0 (1):\penalty0 10783, 2024.
\newblock \doi{https://doi.org/10.1038/s41467-024-55010-6}.

\bibitem[Sutton et~al.(2009)Sutton, Orteu, and Schreier]{Sutton2009}
Michael~A. Sutton, Jean-Jos{\'e} Orteu, and Hubert Schreier.
\newblock \emph{Image correlation for shape, motion and deformation measurements: basic concepts, theory and applications}.
\newblock Springer Science \& Business Media, 2009.

\bibitem[Qi and Harris(2010)]{qi2010in}
Y.~Qi and S.J. Harris.
\newblock In situ observation of strains during lithiation of a graphite electrode.
\newblock \emph{Journal of The Electrochemical Society}, 157\penalty0 (6):\penalty0 A741, 2010.
\newblock \doi{10.1149/1.3377130}.

\bibitem[Jones et~al.(2014)Jones, Silberstein, White, and Sottos]{jones2014situ}
Elias M.~C. Jones, Meredith~N. Silberstein, Scott~R. White, and Nancy~R. Sottos.
\newblock In situ measurements of strains in composite battery electrodes during electrochemical cycling.
\newblock \emph{Experimental Mechanics}, 54\penalty0 (6):\penalty0 971--985, 2014.
\newblock \doi{10.1007/s11340-014-9873-3}.

\bibitem[Xie et~al.(2021)Xie, Han, Song, Li, Kang, and Zhang]{xie2021insitu}
Hui Xie, Bing Han, Haowei Song, Xiaoyu Li, Yilan Kang, and Qiao Zhang.
\newblock In-situ measurements of electrochemical stress/strain fields and stress analysis during an electrochemical process.
\newblock \emph{Journal of the Mechanics and Physics of Solids}, 156:\penalty0 104602, 2021.
\newblock \doi{https://doi.org/10.1016/j.jmps.2021.104602}.

\bibitem[Zhao et~al.(2020{\natexlab{a}})Zhao, Storey, Braatz, and Bazant]{Zhao2020}
H.~Zhao, B.~D. Storey, R.~D. Braatz, and M.~Z. Bazant.
\newblock Learning the physics of pattern formation from images.
\newblock \emph{Physical Review Letters}, 124\penalty0 (6):\penalty0 060201, 2020{\natexlab{a}}.
\newblock \doi{10.1103/PhysRevLett.124.060201}.

\bibitem[Zhao et~al.(2021)Zhao, Braatz, and Bazant]{Zhao2021}
H.~Zhao, R.~D. Braatz, and M.~Z. Bazant.
\newblock Image inversion and uncertainty quantification for constitutive laws of pattern formation.
\newblock \emph{Journal of Computational Physics}, 436:\penalty0 110279, 2021.
\newblock \doi{10.1016/j.jcp.2021.110279}.

\bibitem[Zhao et~al.(2023{\natexlab{a}})Zhao, Deng, Cohen, Lim, Li, Fraggedakis, Jiang, Storey, Chueh, Braatz, and Bazant]{Zhao2023}
H.~Zhao, H.~D. Deng, A.~E. Cohen, J.~Lim, Y.~Li, D.~Fraggedakis, B.~Jiang, B.~D. Storey, W.~C. Chueh, R.~D. Braatz, and M.~Z. Bazant.
\newblock Learning heterogeneous reaction kinetics from x-ray videos pixel by pixel.
\newblock \emph{Nature}, 621\penalty0 (7978):\penalty0 289--294, 2023{\natexlab{a}}.
\newblock \doi{https://doi.org/10.1038/s41586-023-06393-x}.

\bibitem[Akerson et~al.(2025)Akerson, Rajan, and Bhattacharya]{akerson2025learning}
A.~Akerson, A.~Rajan, and K.~Bhattacharya.
\newblock Learning constitutive relations from experiments: 1. pde constrained optimization.
\newblock \emph{Journal of the Mechanics and Physics of Solids}, page 106128, 2025.
\newblock \doi{https://doi.org/10.1016/j.jmps.2025.106128}.

\bibitem[Raissi et~al.(2019)Raissi, Perdikaris, and Karniadakis]{Raissi2019}
Maziar Raissi, Paris Perdikaris, and George~Em Karniadakis.
\newblock Physics-informed neural networks: A deep learning framework for solving forward and inverse problems involving nonlinear partial differential equations.
\newblock \emph{Journal of Computational Physics}, 378:\penalty0 686--707, 2019.
\newblock \doi{https://doi.org/10.1016/j.jcp.2018.10.045}.

\bibitem[Karniadakis et~al.(2021)Karniadakis, Kevrekidis, Lu, Perdikaris, Wang, and Yang]{karniadakis2021physics}
George~E Karniadakis, Ioannis~G Kevrekidis, Lu~Lu, Paris Perdikaris, Sifan Wang, and Liu Yang.
\newblock Physics-informed machine learning.
\newblock \emph{Nature Reviews Physics}, 3\penalty0 (6):\penalty0 422--440, 2021.
\newblock \doi{https://doi.org/10.1038/s42254-021-00314-5}.

\bibitem[Cottrill et~al.(2020)Cottrill, Goulet, Fremy, Meulemans, Sheldon, and Bazant]{cottrill2020simultaneous}
A.L. Cottrill, R.~Goulet, F.~Fremy, J.~Meulemans, M.R. Sheldon, and M.Z. Bazant.
\newblock Simultaneous inversion of optical and infra-red image data to determine thermo-mechanical properties of thermally conductive solid materials.
\newblock \emph{International Journal of Heat and Mass Transfer}, 163:\penalty0 120445, 2020.
\newblock \doi{10.1016/j.ijheatmasstransfer.2020.120445}.

\bibitem[Ulvestad et~al.(2015)Ulvestad, Singer, Clark, Cho, Kim, Harder, Maser, Meng, and Shpyrko]{ulvestad2015topological}
A~Ulvestad, A~Singer, JN~Clark, HM~Cho, JW~Kim, R~Harder, J~Maser, YS~Meng, and OG~Shpyrko.
\newblock Topological defect dynamics in operando battery nanoparticles.
\newblock \emph{Science}, 348\penalty0 (6241):\penalty0 1344--1347, 2015.
\newblock \doi{10.1126/science.aaa1313}.

\bibitem[Ihuaenyi et~al.(2024)Ihuaenyi, Luo, Li, and Zhu]{ihuaenyi2024seeking}
RC~Ihuaenyi, J~Luo, W~Li, and J~Zhu.
\newblock Seeking the most informative design of test specimens for learning constitutive models.
\newblock \emph{Extreme Mechanics Letters}, 69:\penalty0 102169, 2024.
\newblock \doi{https://doi.org/10.1016/j.eml.2024.102169}.

\bibitem[Ihuaenyi et~al.(2025{\natexlab{a}})Ihuaenyi, Li, Bazant, and Zhu]{ihuaenyi2025mechanics}
R.C. Ihuaenyi, W.~Li, M.Z. Bazant, and J.~Zhu.
\newblock Mechanics informatics: A paradigm for efficiently learning constitutive models.
\newblock \emph{Journal of the Mechanics and Physics of Solids}, page 106239, 2025{\natexlab{a}}.
\newblock \doi{https://doi.org/10.1016/j.jmps.2025.106239}.

\bibitem[Cahn and Hilliard(1958)]{Cahn1958}
J.~W. Cahn and J.~E. Hilliard.
\newblock Free energy of a nonuniform system. i. interfacial free energy.
\newblock \emph{The Journal of Chemical Physics}, 28\penalty0 (2):\penalty0 258--267, 1958.
\newblock \doi{10.1063/1.1744102}.

\bibitem[Cahn(1965)]{Cahn1965}
J.~W. Cahn.
\newblock Phase separation by spinodal decomposition in isotropic systems.
\newblock \emph{The Journal of Chemical Physics}, 42\penalty0 (1):\penalty0 93--99, 1965.
\newblock \doi{https://doi.org/10.1063/1.1695731}.

\bibitem[Lim et~al.(2016)Lim, Li, Lun, Li, Zhong, Jian, Huang, Chen, Yu, Xin, and et~al.]{Lim2016}
J.~Lim, Y.~Li, Z.~Lun, H.~Li, L.~Zhong, Z.~Jian, J.~Huang, G.~Chen, Y.~Yu, H.~L. Xin, and et~al.
\newblock Origin and hysteresis of lithium compositional spatiodynamics within battery primary particles.
\newblock \emph{Science}, 353\penalty0 (6299):\penalty0 566--571, 2016.
\newblock \doi{10.1126/science.aaf4914}.

\bibitem[Mefford et~al.(2021)Mefford, Kurilovich, Saunders, Hardin, Abakumov, Forslund, Bonnefont, Dai, Johnston, and Stevenson]{Mefford2021}
J.~T. Mefford, A.~A. Kurilovich, J.~Saunders, W.~G. Hardin, A.~M. Abakumov, R.~P. Forslund, A.~Bonnefont, S.~Dai, K.~P. Johnston, and K.~J. Stevenson.
\newblock Correlative operando microscopy of oxygen evolution electrocatalysts.
\newblock \emph{Nature}, 593:\penalty0 67--73, 2021.
\newblock \doi{10.1038/s41586-021-03454-x}.

\bibitem[Bazant(2013)]{Bazant2013}
M.~Z. Bazant.
\newblock Theory of chemical kinetics and charge transfer based on nonequilibrium thermodynamics.
\newblock \emph{Accounts of Chemical Research}, 46\penalty0 (5):\penalty0 1144--1160, 2013.
\newblock \doi{10.1021/ar300145c}.

\bibitem[Bazant(2017)]{Bazant2017}
M.~Z. Bazant.
\newblock Thermodynamic stability of driven open systems and control of phase separation by electro-autocatalysis.
\newblock \emph{Faraday Discussions}, 199:\penalty0 423--463, 2017.
\newblock \doi{DOI https://doi.org/10.1039/C7FD00037E}.

\bibitem[Fraggedakis et~al.(2021)Fraggedakis, McEldrew, Smith, Krishnan, Zhang, Bai, Chueh, Shao-Horn, and Bazant]{Fraggedakis2021}
Dionisios Fraggedakis, Matthew McEldrew, Robert~B. Smith, Yashaswini Krishnan, Yiyang Zhang, Peng Bai, William~C. Chueh, Yang Shao-Horn, and Martin~Z. Bazant.
\newblock Theory of coupled ion-electron transfer kinetics.
\newblock \emph{Electrochimica Acta}, 367:\penalty0 137432, 2021.
\newblock \doi{10.1016/j.electacta.2020.137432}.

\bibitem[Doyle et~al.(1993)Doyle, Fuller, and Newman]{doyle1993modeling}
M.~Doyle, T.~F. Fuller, and J.~Newman.
\newblock Modeling of galvanostatic charge and discharge of the lithium/polymer/insertion cell.
\newblock \emph{Journal of the Electrochemical Society}, 140\penalty0 (6):\penalty0 1526--1533, 1993.
\newblock \doi{10.1149/1.2221597}.

\bibitem[Cogswell and Bazant(2012)]{cogswell2012coherency}
D.~A. Cogswell and M.~Z. Bazant.
\newblock Coherency strain and the kinetics of phase separation in lifepo4 nanoparticles.
\newblock \emph{ACS Nano}, 6\penalty0 (3):\penalty0 2215--2225, 2012.
\newblock \doi{10.1021/nn204177u}.

\bibitem[Ihuaenyi et~al.(2025{\natexlab{b}})Ihuaenyi, Fang, Ashok, Condon, Jiao, Attia, Li, and Zhu]{Ihuaenyi2025}
R.C. Ihuaenyi, R.~Fang, A.S. Ashok, A.~Condon, J.~Jiao, P.M. Attia, W.~Li, and J.~Zhu.
\newblock Lifetime extension of aged li-ion prismatic batteries via mechanical constraints.
\newblock \emph{Cell Reports Physical Science}, 6\penalty0 (7), 2025{\natexlab{b}}.
\newblock \doi{https://doi.org/10.1016/j.xcrp.2025.102685}.

\bibitem[Zhang et~al.(2021)Zhang, Yang, Ren, Wang, and He]{Zhang2021}
Huan Zhang, Ying Yang, Danyang Ren, Lin Wang, and Xiangming He.
\newblock Graphite as anode materials: Fundamental mechanism, recent progress and advances.
\newblock \emph{Energy Storage Materials}, 36:\penalty0 147--170, 2021.
\newblock \doi{https://doi.org/10.1016/j.ensm.2020.12.027}.

\bibitem[Martínez-Alonso et~al.(2022)Martínez-Alonso, Guevara-Vela, and LLorca]{MartinezAlonso2022}
C.~Martínez-Alonso, J.~M. Guevara-Vela, and J.~LLorca.
\newblock Understanding the effect of mechanical strains on the catalytic activity of transition metals.
\newblock \emph{Physical Chemistry Chemical Physics}, 24\penalty0 (8):\penalty0 4832--4842, 2022.
\newblock \doi{DOI https://doi.org/10.1039/D1CP05436H}.

\bibitem[Hua et~al.(2021)Hua, Kim, Du, Wu, Bai, and He]{hua2021swaying}
Mingchao Hua, Changjin Kim, Yunlong Du, Dayong Wu, Ruobing Bai, and Xuanhe He.
\newblock Swaying gel: Chemo-mechanical self-oscillation based on dynamic buckling.
\newblock \emph{Matter}, 4\penalty0 (3):\penalty0 1029--1041, 2021.
\newblock \doi{https://doi.org/10.1016/j.matt.2021.01.002}.

\bibitem[Ihuaenyi et~al.(2021)Ihuaenyi, Yan, Deng, Bae, Sakib, and Xiao]{ihuaenyi2021orthotropic}
R.~C. Ihuaenyi, S.~Yan, J.~Deng, C.~Bae, I.~Sakib, and X.~Xiao.
\newblock Orthotropic thermo-viscoelastic model for polymeric battery separators with electrolyte effect.
\newblock \emph{Journal of The Electrochemical Society}, 168\penalty0 (9):\penalty0 090536, 2021.
\newblock \doi{https://doi.org/10.1149/1945-7111/ac24b6}.

\bibitem[Girard et~al.(2007)Girard, Cavalcanti-Adam, Kemkemer, and Spatz]{girard2007cellular}
Philippe~P. Girard, Elisabetta~A. Cavalcanti-Adam, Ralf Kemkemer, and Joachim~P. Spatz.
\newblock Cellular chemomechanics at interfaces: sensing, integration and response.
\newblock \emph{Soft Matter}, 3\penalty0 (3):\penalty0 307--326, 2007.
\newblock \doi{10.1039/b614008d}.

\bibitem[Bai and Bazant(2014)]{Bai2014}
P.~Bai and M.~Z. Bazant.
\newblock Charge transfer kinetics at the solid–solid interface in porous electrodes.
\newblock \emph{Nature Communications}, 5:\penalty0 3585, 2014.
\newblock \doi{10.1038/ncomms4585}.

\bibitem[Callen(1985)]{Callen1985}
H.~B. Callen.
\newblock \emph{Thermodynamics and an Introduction to Thermostatistics}.
\newblock John Wiley \& Sons, New York, 1985.

\bibitem[Qi et~al.(2010)Qi, Guo, Hector, and Timmons]{Qi2010}
Y.~Qi, H.~Guo, L.~G. Hector, and A.~Timmons.
\newblock Threefold increase in the young’s modulus of graphite negative electrode during lithium intercalation.
\newblock \emph{Journal of The Electrochemical Society}, 157\penalty0 (5):\penalty0 A558, 2010.
\newblock \doi{10.1149/1.3327913}.

\bibitem[Kganyago and Ngoepe(2003)]{Kganyago2003}
K.~R. Kganyago and P.~E. Ngoepe.
\newblock Structural and electronic properties of lithium intercalated graphite lic6.
\newblock \emph{Physical Review B}, 68\penalty0 (20):\penalty0 205111, 2003.
\newblock \doi{10.1103/PhysRevB.68.205111}.

\bibitem[Blakslee et~al.(1970)Blakslee, Proctor, Seldin, Spence, and Weng]{Blakslee1970}
O.~L. Blakslee, D.~G. Proctor, E.~J. Seldin, G.~B. Spence, and T.~Weng.
\newblock Elastic constants of compression‐annealed pyrolytic graphite.
\newblock \emph{Journal of Applied Physics}, 41\penalty0 (8):\penalty0 3373--3382, 1970.
\newblock \doi{https://doi.org/10.1063/1.1659428}.

\bibitem[Doyle et~al.(1996)Doyle, Newman, Gozdz, Schmutz, and Tarascon]{doyle1996comparison}
M.~Doyle, J.~Newman, A.~S. Gozdz, C.~N. Schmutz, and J.~M. Tarascon.
\newblock Comparison of modeling predictions with experimental data from plastic lithium ion cells.
\newblock \emph{Journal of the Electrochemical Society}, 143\penalty0 (6):\penalty0 1890, 1996.
\newblock \doi{10.1149/1.1836921}.

\bibitem[Taghikhani et~al.(2020)Taghikhani, Weddle, Berger, and Kee]{taghikhani2020chemo}
K.~Taghikhani, P.J. Weddle, J.R. Berger, and R.J. Kee.
\newblock Chemo-mechanical behavior of highly anisotropic and isotropic polycrystalline graphite particles during lithium intercalation.
\newblock \emph{Journal of The Electrochemical Society}, 167\penalty0 (11):\penalty0 110554, 2020.
\newblock \doi{10.1149/1945-7111/aba5d2}.

\bibitem[Persson et~al.(2010{\natexlab{a}})Persson, Sethuraman, Hardwick, Hinuma, Meng, van~der Ven, Srinivasan, Kostecki, and Ceder]{persson2010lithium}
K.~Persson, V.~A. Sethuraman, L.~J. Hardwick, Y.~Hinuma, Y.~S. Meng, A.~van~der Ven, V.~Srinivasan, R.~Kostecki, and G.~Ceder.
\newblock Lithium diffusion in graphitic carbon.
\newblock \emph{J. Phys. Chem. Lett.}, 1\penalty0 (8):\penalty0 1176--1181, 2010{\natexlab{a}}.
\newblock \doi{https://doi.org/10.1021/jz100188d}.

\bibitem[Persson et~al.(2010{\natexlab{b}})Persson, Hinuma, Meng, van~der Ven, and Ceder]{persson2010thermodynamic}
K.~Persson, Y.~Hinuma, Y.~S. Meng, A.~van~der Ven, and G.~Ceder.
\newblock Thermodynamic and kinetic properties of the li-graphite system from first-principles calculations.
\newblock \emph{Phys. Rev. B}, 82\penalty0 (12):\penalty0 125416, 2010{\natexlab{b}}.
\newblock \doi{10.1103/PhysRevB.82.125416}.

\bibitem[Lian and Bazant(2024)]{Lian2024}
H.~Lian and M.~Z. Bazant.
\newblock Modeling lithium plating onset on porous graphite electrodes under fast charging with hierarchical multiphase porous electrode theory.
\newblock \emph{Journal of The Electrochemical Society}, 171\penalty0 (1):\penalty0 010526, 2024.
\newblock \doi{10.1149/1945-7111/ad33f8}.

\bibitem[Maxisch and Ceder(2006)]{maxisch2006elastic}
T.~Maxisch and G.~Ceder.
\newblock Elastic properties of olivine li$_x$fepo$_4$ from first principles.
\newblock \emph{Phys. Rev. B}, 73\penalty0 (17):\penalty0 174112, 2006.
\newblock \doi{10.1103/PhysRevB.73.174112}.

\bibitem[Ouyang et~al.(2004)Ouyang, Shi, Wang, Huang, and Chen]{ouyang2004}
C.~Ouyang, S.~Shi, Z.~Wang, X.~Huang, and L.~Chen.
\newblock First-principles study of li ion diffusion in \( \text{LiFePO}_4 \).
\newblock \emph{Physical Review B}, 69\penalty0 (10):\penalty0 104303, 2004.
\newblock \doi{10.1103/PhysRevB.69.104303}.

\bibitem[Morgan et~al.(2003)Morgan, Van~der Ven, and Ceder]{Morgan2003}
D.~Morgan, A.~Van~der Ven, and G.~Ceder.
\newblock Li conductivity in lixmpo$_4$ (m = mn, fe, co, ni) olivine materials.
\newblock \emph{Electrochem. Solid-State Lett.}, 7\penalty0 (2):\penalty0 A30, 2003.
\newblock \doi{10.1149/1.1633511}.

\bibitem[Malik et~al.(2010)Malik, Burch, Bazant, and Ceder]{Malik2010}
R.~Malik, D.~Burch, M.~Bazant, and G.~Ceder.
\newblock Particle size dependence of the ionic diffusivity.
\newblock \emph{Nano Letters}, 10\penalty0 (10):\penalty0 4123--4127, 2010.
\newblock \doi{10.1021/nl1023595}.

\bibitem[Nishimura et~al.(2008)Nishimura, Kobayashi, Ohoyama, Kanno, Yashima, and Yamada]{nishimura2008}
Shin-ichi Nishimura, Genki Kobayashi, Kenji Ohoyama, Ryoji Kanno, Masatomo Yashima, and Atsuo Yamada.
\newblock Experimental visualization of lithium diffusion in \( \text{Li}_x\text{FePO}_4 \).
\newblock \emph{Nature Materials}, 7:\penalty0 707--711, 2008.
\newblock \doi{10.1038/nmat2251}.

\bibitem[Zhang(2011)]{Zhang2011}
W.~J. Zhang.
\newblock Structure and performance of lifepo$_4$ cathode materials: A review.
\newblock \emph{Journal of Power Sources}, 196\penalty0 (6):\penalty0 2962--2970, 2011.
\newblock \doi{https://doi.org/10.1016/j.jpowsour.2010.11.113}.

\bibitem[Zhao et~al.(2020{\natexlab{b}})Zhao, Storey, Braatz, and Bazant]{zhao2020learning}
H.~Zhao, B.~D. Storey, R.~D. Braatz, and M.~Z. Bazant.
\newblock Learning the physics of pattern formation from images.
\newblock \emph{Physical Review Letters}, 124\penalty0 (6):\penalty0 060201, 2020{\natexlab{b}}.
\newblock \doi{10.1103/PhysRevLett.124.060201}.

\bibitem[Zhao et~al.(2023{\natexlab{b}})Zhao, Deng, Cohen, Lim, Li, Fraggedakis, Jiang, Storey, Chueh, Braatz, and Bazant]{zhao2023learning}
H.~Zhao, H.~D. Deng, A.~E. Cohen, J.~Lim, Y.~Li, D.~Fraggedakis, B.~Jiang, B.~D. Storey, W.~C. Chueh, R.~D. Braatz, and M.~Z. Bazant.
\newblock Learning heterogeneous reaction kinetics from x-ray videos pixel by pixel.
\newblock \emph{Nature}, 621\penalty0 (7978):\penalty0 289--294, 2023{\natexlab{b}}.
\newblock \doi{https://doi.org/10.1038/s41586-023-06393-x}.

\bibitem[Bazant(2023)]{Bazant2023}
Martin~Z. Bazant.
\newblock Unified quantum theory of electrochemical kinetics by coupled ion--electron transfer.
\newblock \emph{Faraday Discussions}, 246:\penalty0 60--124, 2023.
\newblock \doi{https://doi.org/10.1039/D3FD00108C}.

\bibitem[Wang(2008)]{wang2008karhunen}
L.~Wang.
\newblock \emph{Karhunen-Loeve expansions and their applications}.
\newblock PhD thesis, London School of Economics and Political Science (United Kingdom), 2008.

\bibitem[Mockus et~al.(1978)Mockus, Tiesis, and Zilinskas]{mockus1978application}
Jonas Mockus, Vytautas Tiesis, and Antanas Zilinskas.
\newblock The application of bayesian methods for seeking the extremum.
\newblock \emph{Towards global optimization}, 2\penalty0 (117-129):\penalty0 2, 1978.

\bibitem[Jones et~al.(1998)Jones, Schonlau, and Welch]{jones1998efficient}
Donald~R Jones, Matthias Schonlau, and William~J Welch.
\newblock Efficient global optimization of expensive black-box functions.
\newblock \emph{Journal of Global optimization}, 13\penalty0 (4):\penalty0 455--492, 1998.

\bibitem[Bergstra et~al.(2011)Bergstra, Bardenet, Bengio, and K{\'e}gl]{bergstra2011algorithms}
James Bergstra, R{\'e}mi Bardenet, Yoshua Bengio, and Bal{\'a}zs K{\'e}gl.
\newblock Algorithms for hyper-parameter optimization.
\newblock In \emph{Advances in neural information processing systems}, volume~24, pages 2546--2554, 2011.

\bibitem[Fletcher(2000)]{Fletcher2000}
R.~Fletcher.
\newblock \emph{Practical Methods of Optimization}.
\newblock John Wiley \& Sons, Ltd, Chichester, West Sussex, England, 2000.

\end{thebibliography}
\clearpage
\newpage

\setcounter{figure}{0}
\setcounter{equation}{0}
\renewcommand{\thefigure}{S\arabic{figure}}
\renewcommand{\theequation}{S\arabic{equation}}

\begin{center}
{\Large{\bf Supporting Information}}\\[0.3cm]
\end{center}

\section*{Supplementary Note 1: Model parametrization}

In this study, the known parameter values for the investigated material systems are derived from literature values obtained through experimental measurements or first-principles calculations. These parameters serve as benchmarks for understanding and modeling the chemomechanical phenomena. We focus on two material systems, each with its underlying physics explicitly embedded within the studied domain. The first is a graphite intercalation compound (\( \text{Li}_x\text{C}_6 \)) and the second is a lithium iron phosphate platelet (\( \text{Li}_x\text{FePO}_4 \)).

\subsection*{Graphite system}

\subsubsection*{Anisotropic stiffness tensor}

The components of the anisotropic stiffness tensor and the material density define the mechanical properties of graphite used in the simulations. While \( \text{Li}_x\text{C}_6 \) crystals exhibit six distinct stiffness coefficients, the elastic constant tensor \( C_{ij} \) contains only five independent constants, in contrast to the 21 typically required for fully anisotropic materials. The stiffness tensor for graphite is expressed as:

\begin{equation}
C_{ij} =
\begin{bmatrix}
C_{11} & C_{12} & C_{13} & 0 & 0 & 0 \\
C_{12} & C_{11} & C_{13} & 0 & 0 & 0 \\
C_{13} & C_{13} & C_{33} & 0 & 0 & 0 \\
0 & 0 & 0 & C_{44} & 0 & 0 \\
0 & 0 & 0 & 0 & C_{44} & 0 \\
0 & 0 & 0 & 0 & 0 & C_{66}
\end{bmatrix},
\end{equation}
where \( C_{66} = \frac{1}{2}(C_{11} - C_{12}) \). These elastic constants quantify the crystalline bonding strength or stiffness in specific crystallographic directions. Here, \( C_{11} \) represents the stiffness within the AB plane (Fig. \ref{fig: Figure S7}A), \( C_{33} \) corresponds to the interlayer stiffness along the \( c \)-axis, and \( C_{44} \) and \( C_{66} \) are the shear moduli. The combination \( C_{11} + C_{12} \) defines the intralayer stiffness parallel to the AB plane, while \( C_{12} \) and \( C_{13} \) are mixed-index moduli, with \( C_{13} \) capturing the stiffness in a linear combination of the \( c \)-axis direction and directions parallel to the AB plane.

Although the stiffness tensor for graphite and lithiated graphite has been derived from first-principles calculations using density functional theory (DFT) \cite{Qi2010, Kganyago2003}, we employ experimentally determined values \cite{Blakslee1970}, with $C_{11}=1060$, $C_{33}=36.5$, $C_{44}=4.5$, $C_{66}=440$, $C_{12}=180$, and $C_{13}=15$. The unit of the stiffness components is MPa. Additionally, the density of graphite is set as \( 1900 \, \text{kg/m}^3 \) \cite{doyle1996comparison}.

\subsubsection*{Chemical expansion}

Chemical expansion in graphite is inherently anisotropic, with distinct dimensional changes occurring within the AB plane (in-plane) and between the graphene layers (out-of-plane)(Fig. \ref{fig: Figure S6}a). Considering the a-c plane, the anisotropic chemical expansion can be described as a diagonal tensor with components \(\beta_{a}\) and \(\beta_{c}\).  \(\beta_{a}\) represents the chemical expansion coefficient along the \(a\)-axis (in-plane), and \(\beta_{c}\) corresponds to the chemical expansion coefficient along the \(c\)-axis (out-of-plane). 

The anisotropic chemical expansion coefficients for the stable equilibrium phases of graphite were analytically derived by Taghikhani et al. \cite{taghikhani2020chemo}. These coefficients are expressed as step functions of the lithium-ion intercalation fraction (or concentration), as illustrated in Fig. \ref{fig: Figure S13}. These step functions are explicitly incorporated into the model to accurately describe the phase-dependent lattice response. 
 
\subsubsection*{Diffusivity}

Lithium diffusion within the graphite basal plane occurs at a much higher rate compared to diffusion perpendicular to the planes~\cite{persson2010lithium, persson2010thermodynamic}. For example, first principle calculations by Persson et al. reported an average diffusion coefficient of approximately \(10^{-10} \, \text{m}^2\text{s}^{-1}\) for in-plane diffusion and of approximately \(10^{-16} \, \text{m}^2\text{s}^{-1}\) for out-of-plane diffusion~\cite{persson2010lithium}. Additionally, the diffusion coefficient can vary depending on the phase and, consequently, the lithium concentration. A first-principles study suggested that the in-plane diffusion coefficients are concentration-dependent~\cite{persson2010thermodynamic}. Figure~\ref{fig: Figure S14} presents the transverse, concentration-dependent diffusivity profile \(D_a\) from the Persson et al. study (red dots) ~\cite{persson2010thermodynamic}, an interpolation of Persson's results by Lian et al \cite{Lian2024} which accounts for phase transitions and the corresponding fit used in the current model. The out-of-plane diffusion coefficient \(D_c\) is assumed to remain constant at \(8.7 \times 10^{-16} \, \text{m}^2\text{s}^{-1}\)~\cite{persson2010lithium}. It is important to note that \(D_c\) is six orders of magnitude smaller than \(D_a\), hence, the dominant diffusivity profile (\(D_a\)) is inferred in this work. 

\subsection*{Lithium iron phosphate system}

\subsubsection*{Anisotropic stiffness tensor}

The olivine-type (LFP) crystal exhibits pronounced structural anisotropy due to its orthorhombic symmetry, which is reflected in its mechanical properties. The crystal structure consists of corner-sharing \texorpdfstring{FeO$_6$}{FeO6} octahedra forming chains along the [010] direction, with \texorpdfstring{PO$_4$}{PO4} tetrahedra linking these chains and creating channels for lithium-ion diffusion along the [010] axis \cite{maxisch2006elastic}. This anisotropic arrangement leads to significant variations in mechanical stiffness along different crystallographic directions.

The stiffness tensor components for \texorpdfstring{FePO$_4$}{FePO4} used in this work were derived from first-principles calculations employing the generalized gradient approximation method with Hubbard correction (GGA+U) \cite{maxisch2006elastic}. The full stiffness tensor takes the form:

\begin{equation}
C_{ij} =
\begin{bmatrix}
C_{11} & C_{12} & C_{13} & 0 & 0 & 0 \\
C_{12} & C_{22} & C_{23} & 0 & 0 & 0 \\
C_{13} & C_{23} & C_{33} & 0 & 0 & 0 \\
0 & 0 & 0 & C_{44} & 0 & 0 \\
0 & 0 & 0 & 0 & C_{55} & 0 \\
0 & 0 & 0 & 0 & 0 & C_{66}
\end{bmatrix},
\end{equation}

with $C_{11}=175.9$, $C_{22}=153.6$, $C_{33}=135.0$, $C_{44}=38.8$, $C_{55}=47.5$, $C_{66}=55.6$, $C_{12}=29.6$, $C_{13}=54.0$, and $C_{23}=19.6$. The unit of the stiffness components is also MPa.

This structural anisotropy has important implications for lithium intercalation dynamics, as the different stiffness values along principal directions affect both stress development during cycling. Additionally, the density of \texorpdfstring{LiFePO$_4$}{LiFePO4} is set as 3,490 kg/m$^3$ \cite{maxisch2006elastic}, consistent with its tightly packed olivine structure.

\subsubsection*{Diffusivity}

Lithium diffusivity in \( \text{LiFePO}_4 \) (LFP) crystals is highly anisotropic due to the olivine crystal structure (Fig. \ref{fig: Figure S6}b), which features one-dimensional channels along the [010] direction (b-axis). These channels provide a low-energy pathway for lithium-ion migration, resulting in significantly faster diffusion along the [010] direction compared to the [100] and [001] perpendicular directions \cite{ouyang2004, Morgan2003, Malik2010}. The anisotropy arises from structural constraints and higher energy barriers for lithium movement in directions other than the [010] axis. In this model, the effective diffusivity is taken to be that in the dominant [010] direction. This approach simplifies the computational framework while maintaining physical relevance, as the contribution of diffusion in other directions is negligible due to their orders-of-magnitude lower diffusivity. This simplification is consistent with experimental and theoretical studies that highlight the one-dimensional nature of lithium diffusion in LFP \cite{ouyang2004, Malik2010, nishimura2008}. 

Although effective ionic diffusivity depends on factors such as the concentration of obstructing defects in the channel and the particle size \cite{Malik2010}, here we consider its dependence on ionic concentration. The concentration-dependent diffusivity of lithium in \( \text{LiFePO}_4 \) is nonlinear mainly due to phase separation and electrochemical potential effects \cite{Zhang2011}. Moreover, Morgan et al. \cite{Morgan2003} determined the diffusivity of lithium along 1D chains in \( \text{LiFePO}_4 \) using first-principles calculations as \( 10^{-7} \, \text{cm}^2/\text{s} \) at a concentration fraction of 0 and \( 10^{-8} \, \text{cm}^2/\text{s} \) at a concentration fraction of 1. Hence, we represent the concentration-dependent diffusivity profile as a linear function that spans the values determined by Morgan et al \cite{Morgan2003}, as shown in Figure \ref{fig: Figure S15}.

\subsubsection*{Chemical expansion}

Chemical expansion in lithium iron phosphate exhibits anisotropy, as demonstrated in a previous study \cite{deng2022correlative}. This anisotropic behaviour is also described as a diagonal tensor considering the a-c plane. With $\beta_{a}$ and $\beta_{c}$ representing the chemical expansion coefficients along the crystallographic $a$- and $c$-axes, respectively. We restrict our consideration to the two-dimensional projection within the a-c crystallographic plane, as that coincides with the minor and major axes of the scanning transmission X-ray microscopy (STXM) images of the particle. In this work, the chemical strain in the a-c plane is approximated by a linear dependence on Li concentration, based on the result inverted from ptychography and the
strain map \cite{deng2022correlative}.

\subsubsection*{Free energy and exchange current}

We describe the homogeneous chemical free energy using the regular solution model:
\begin{equation}
    \frac{g_h(\bar{c})}{k_BT} = \bar{c} \ln \bar{c} + (1 - \bar{c}) \ln (1 - \bar{c}) + \Omega \bar{c} (1 - \bar{c}),
    \label{eq:free_energy}
\end{equation}  
where $\Omega$ represents the interaction parameter that governs phase behavior and $\bar{c}$ denotes the normalized concentration. Positive values of $\Omega$ promote phase separation, while negative values favor mixing. Following established literature~\cite{cogswell2012coherency}, $\Omega = 4.47$, which corresponds to a miscibility gap with spinodal points at $\bar{c}_1 = 0.0126$ and $\bar{c}_2 = 1 - \bar{c}_1$ due to symmetry. Other values taken from literature include, the gradient energy coefficient, $\kappa = 5.02 \times 10^{-10}$J/m and maximum concentration $c_{max}=2.29 \times 10^{4}$ mol/$\text{m}^3$ ~\cite{cogswell2012coherency}. Also, the absolute temperature value used in non-dimensionalization is $T=298 K$.

The homogeneous chemical potential, derived from $\partial g_h/\partial c$ and non-dimensionalized by $k_BT$, takes the form:
\begin{equation}
    \mu_h(\bar{c}) = \ln \left( \frac{c}{1 - \bar{c}} \right) + \Omega (1 - 2\bar{c}).
    \label{eq:chemical_potential_regular}
\end{equation}

In the inverse learning framework, we employ a physically constrained representation of the chemical potential that combines an ideal entropic contribution with a polynomial expansion capturing excess effects. This ensures the concentration remains within the physical bounds $[0, 1]$:
\begin{equation}
\mu_h(\bar{c}) = \ln \left( \frac{\bar{c}}{1 - \bar{c}} \right) + \sum_{i=1}^{M} a_i P_i(\bar{c}),
\label{eq:chemical_potential_learned}
\end{equation}
where $P_i(\bar{c})$ are Legendre polynomials defined on $[0, 1]$ and $\{a_i\}$ are the coefficients to be learned. This formulation maintains consistency with the regular solution framework~\cite{zhao2020learning,zhao2023learning} while providing sufficient flexibility for parameter identification.

For benchmark verification of the inversion framework, we utilize the known case where $\Omega = 3$~\cite{zhao2020learning} to test the case study of learning pattern formation from mechanical data. In contrast, when learning heterogeneous reaction kinetics from experimental images, the free energy landscape remains unknown \textit{a priori} and is determined through the PDE-constrained optimization procedure.

It is noteworthy that the chemical potential is constrained to satisfy thermodynamic equilibrium conditions at the miscibility gap boundaries:
\begin{align}
\int_{\bar{c}_1}^{\bar{c}_2} \mu_h(\bar{c})\, d\bar{c} &= \mu_h(\bar{c}_1)(\bar{c}_2 - \bar{c}_1), \label{eq:maxwell_construction} \\
\mu_h(\bar{c}_1) &= \mu_h(\bar{c}_2). \label{eq:equal_chemical_potential}
\end{align}

The exchange current density requires careful parameterization to ensure physical behavior. Since electron transfer necessitates lattice vacancies~\cite{Fraggedakis2021}, $j_0$ must vanish as $\bar{c} \to 0$ or $\bar{c} \to 1$. We therefore adopt:
\begin{equation}
j_0(\bar{c}) = \bar{c}(1-\bar{c})\sum_{n=1}^{M} b_i P_i(\bar{c}),
\label{eq:exchange_current_learned}
\end{equation}
where $\{b_i\}$ are learnable coefficients and the prefactor $\bar{c}(1-\bar{c})$ enforces the requisite boundary conditions.

We compare our identified exchange current against two established theoretical forms. The first derives from ion-coupled electron transfer (ICET) theory~\cite{Fraggedakis2021,Bazant2023,zhao2023learning}, an extension of coupled-ion-electron transfer theory to the ion-transfer-limited regime:
\begin{equation}
j_0^{\text{ICET}}(\bar{c}) = \bar{c}^{0.66}(1-\bar{c}).
\label{eq:icet_exchange}
\end{equation}

The second represents the empirical form widely employed in porous electrode theory for lithium intercalation with Butler-Volmer kinetics~\cite{doyle1993modeling}:
\begin{equation}
j_0^{\text{PET}}(\bar{c}) = \sqrt{\bar{c}(1-\bar{c})}.
\label{eq:pet_exchange}
\end{equation}

These comparisons provide crucial validation of our learned exchange current functional form against established theoretical predictions and empirical correlations.

\section*{Supplementary Note 2: Spatial heterogeneity}

To model the intrinsic spatial heterogeneity in electrochemical reaction kinetics, we introduce a spatially varying multiplicative prefactor $k(\mathbf{x})$ into the Allen-Cahn equation:
\begin{equation}
\label{eq:allen_cahn}
\frac{\partial c}{\partial t} = k(\mathbf{x}) j_0(c) \left[ \exp\left( - \alpha \tilde{\eta} \right) - \exp\left( (1-\alpha )\tilde{\eta} \right) \right],
\end{equation}
where $\mathbf{x} = (x, y) \in \Omega \subset \mathbb{R}^2$ denotes the spatial coordinate. 

\subsubsection*{Karhunen-Loève expansion of a Gaussian random field}
The heterogeneous field $k(\mathbf{x})$ constitutes one of the fundamental unknowns in the inverse problem, which is learning heterogeneous reaction kinetics from mechanical information. In this work, we model the spatial field as a zero-mean Gaussian process with prescribed statistical structure:
\begin{equation}
k(\mathbf{x}) \sim \mathcal{GP}(\mu(\mathbf{x}), C(\mathbf{x}, \mathbf{x}'))
\end{equation}
where $\mu(\mathbf{x})$ is the mean function (assumed to be zero) and $C(\mathbf{x}, \mathbf{x}')$ is the covariance function encoding spatial correlations.

The theoretical foundation for our approach rests on the Karhunen-Loève Theorem \cite{wang2008karhunen}, which establishes the optimal finite-dimensional representation of stochastic processes in the mean-square sense. For a zero-mean Gaussian process $k(\mathbf{x})$ with covariance function $C(\mathbf{x}, \mathbf{x}')$, the KL expansion provides the decomposition:
\begin{equation}
k(\mathbf{x}) = \sum_{i=1}^{\infty} \sqrt{\lambda_i} \, Z_i \, \phi_i(\mathbf{x}),
\end{equation}
where $\{Z_i\}$ are uncorrelated random variables with $\mathbb{E}[Z_i] = 0$, $\text{Var}[Z_i] = 1$, and $\{\lambda_i, \phi_i(\mathbf{x})\}$ are the eigenvalues and eigenfunctions of the covariance operator $\mathcal{C}$.

The eigenvalue problem is rigorously defined by the Fredholm integral equation of the second kind:
\begin{equation}
\int_{\Omega} C(\mathbf{x}, \mathbf{x}') \phi_i(\mathbf{x}') \, d\mathbf{x}' = \lambda_i \phi_i(\mathbf{x}),
\end{equation}
subject to the orthonormality constraint in $L^2(\Omega)$:
\begin{equation}
\int_{\Omega} \phi_i(\mathbf{x}) \phi_j(\mathbf{x}) \, d\mathbf{x} = \delta_{ij}.
\end{equation}

The eigenvalues satisfy the truncated expansion, which is the fundamental optimality property: 
\begin{equation}
k_M(\mathbf{x}) = \sum_{i=1}^{M} \sqrt{\lambda_i} \, Z_i \, \phi_i(\mathbf{x}).
\end{equation}
The truncated expansion minimizes the mean-square approximation error among all possible $M$-term representations. Through application of the Cauchy-Schwarz inequality and spectral properties of compact self-adjoint operators, the approximation error is rigorously bounded:
\begin{equation}
\mathbb{E}\left[\|k - k_M\|_{L^2(\Omega)}^2\right] = \sum_{i=M+1}^{\infty} \lambda_i
\end{equation}
This bound establishes the eigenvalue decay as a fundamental measure of the field complexity and intrinsic dimensionality.

Furthermore, for computational implementation with discrete observations $\{k(\mathbf{x}_i)\}_{i=1}^N$ at locations $\{\mathbf{x}_i\}_{i=1}^N$, we discretize the infinite-dimensional eigenvalue problem through Galerkin projection onto a finite-dimensional subspace. Let $\mathbf{k} = [k(\mathbf{x}_1), k(\mathbf{x}_2), \ldots, k(\mathbf{x}_N)]^T$ denote the vector of field observations.

The continuous covariance operator $\mathcal{C}$ is approximated by the discrete correlation matrix:
\begin{equation}
\mathbf{C}_{ij} = C(\mathbf{x}_i, \mathbf{x}_j).
\end{equation}

The discrete eigenvalue problem becomes a standard symmetric matrix eigenvalue decomposition:
\begin{equation}
\mathbf{C} \boldsymbol{\phi}_k = \lambda_k \boldsymbol{\phi}_k,
\end{equation}
where $\boldsymbol{\phi}_k = [\phi_k(\mathbf{x}_1), \phi_k(\mathbf{x}_2), \ldots, \phi_k(\mathbf{x}_N)]^T$ are the discrete eigenvectors.

This finite-dimensional discretization preserves the essential spectral properties of the continuous operator in the sense that the eigenvalues remain real and non-negative for symmetric matrices, the eigenvectors maintain orthogonality, and the eigenvalue decay continues to characterize the correlation structure's intrinsic complexity.

\subsubsection*{Spatial correlation learning and field reconstruction framework}

In this work, the heterogeneous field is learned by jointly optimizing the eigenvalue spectrum 
$\{\lambda_k\}$ and spatial correlation parameters directly from the spatial correlation structure inherent in the observed strain field, 
rather than imposing a priori parametric forms. To ensure numerical stability and achieve scale invariance, we standardize the spatial coordinates:
\begin{equation}
\tilde{\mathbf{x}}_i = \frac{\mathbf{x}_i - \boldsymbol{\mu}_{\mathbf{x}}}{\boldsymbol{\sigma}_{\mathbf{x}}},
\end{equation}
where 
\begin{equation}
\boldsymbol{\mu}_{\mathbf{x}} = \frac{1}{N} \sum_{i=1}^N \mathbf{x}_i, 
\qquad 
\boldsymbol{\sigma}_{\mathbf{x}} = \sqrt{\frac{1}{N-1} \sum_{i=1}^N (\mathbf{x}_i - \boldsymbol{\mu}_{\mathbf{x}})^2},
\end{equation}
represent the sample mean and standard deviation, respectively, and $N$ is the number of spatial observation points. 

Using these standardized coordinates, we construct the initial correlation matrix with an exponential kernel:
\begin{equation}
\mathbf{C}^{(0)}_{ij} = \exp\left(-\frac{d_{ij}}{\ell}\right) + \epsilon \delta_{ij},
\end{equation}
where $d_{ij} = \|\tilde{\mathbf{x}}_i - \tilde{\mathbf{x}}_j\|_2$ is the Euclidean distance in standardized 
coordinates, and the regularization parameter $\epsilon > 0$ ensures numerical stability and positive 
definiteness. The correlation length scale $\ell$ is initially determined by nearest-neighbor analysis:
\begin{equation}
\ell^{(0)} = \frac{1}{N} \sum_{i=1}^{N} d_{i,6},
\end{equation}
where $d_{i,6}$ denotes the distance to the 6th nearest neighbor of point $i$, balancing local sensitivity 
with statistical robustness.

Next, we perform eigen-decomposition of the correlation matrix:
\begin{equation}
\mathbf{C}^{(0)} = \boldsymbol{\Phi} \boldsymbol{\Lambda} \boldsymbol{\Phi}^T,
\end{equation}
where $\boldsymbol{\Lambda} = \text{diag}(\lambda_1, \lambda_2, \ldots, \lambda_Q)$ contains eigenvalues ordered by magnitude. 

We retain the $Q$ dominant eigenmodes satisfying the variance retention criterion:
\begin{equation}
\frac{\sum_{k=1}^{Q} \lambda_k}{\sum_{k=1}^{N} \lambda_k} \geq \theta,
\end{equation}
with $\theta = 0.95$, ensuring that the essential correlation structure is captured while reducing dimensionality. 

In this study, the accurate inversion of the heterogeneous reaction kinetics necessitates the 
reconstruction of two distinct spatial fields from the known strain fields. The first is the initial 
concentration field $c_0(\mathbf{x})$ that initializes the forward model, and the second is the 
heterogeneous kinetic prefactor $k(\mathbf{x})$. The reconstruction proceeds through a two-stage process. 
Firstly, the eigenvalue spectrum and initial spatial weights are learned from averaged strain data in a 
data-driven way to capture correlation structures. Secondly, the true spatial weights for both fields 
are learned with physical constraints enforced during the PDE-constrained optimization process.

During the first stage, we establish the connection between the observed strain fields and the unknown 
$c_0(\mathbf{x})$ and $k(\mathbf{x})$ through the assumption that the averaged strain field preserves 
the essential spatial correlation structure of the underlying concentration field through chemomechanical 
coupling. Specifically, we utilize two distinct strain field averages. For the initial concentration field 
$c_0(\mathbf{x})$, we employ the average strain field at the onset of the half-cycle:
\begin{equation}
\bar{\boldsymbol{\varepsilon}}^{(0)} = \boldsymbol{\varepsilon}(\mathbf{x}, t=0).
\end{equation}
For $k(\mathbf{x})$, we use the temporally averaged strain field:
\begin{equation}
\bar{\boldsymbol{\varepsilon}}^{(\text{avg})} = \frac{1}{N_t} \sum_{i=1}^{N_t} \boldsymbol{\varepsilon}(\mathbf{x}, t_i),
\end{equation}
where $N_t$ is the number of time steps, capturing cumulative mechanical signatures of heterogeneous kinetics.

Furthermore, we estimate the spatial weights from the strain field during the training process. 
Given the field observations $\boldsymbol{\varepsilon}(\mathbf{x}, t)$, we solve for the spatial weights 
$\boldsymbol{\alpha} \in \mathbb{R}^N$ . The spatial weights are parameterized to enable efficient optimization while maintaining physical interpretability:
\begin{equation}
\boldsymbol{\alpha}(\boldsymbol{\theta}_{\text{S}}) = \left(s \cdot \boldsymbol{\alpha}^{(0)} + b \cdot \mathbf{1}\right) 
\odot \left(\mathbf{1} + c_1 \boldsymbol{r} + c_2 \boldsymbol{x} + c_3 \boldsymbol{y}\right),
\end{equation}
where $r_i = \|\tilde{\mathbf{x}}_i\|_2$ is the normalized radial distance, $(x_i,y_i)$ are the 
normalized coordinates, and $\boldsymbol{\alpha}^{(0)}$ are baseline weights obtained by solving:
\begin{equation}
(\mathbf{C}^{(0)} + \epsilon \mathbf{I}) \boldsymbol{\alpha}^{(0)} = \bar{\boldsymbol{\varepsilon}},
\end{equation}
where $\epsilon = 10^{-6}$ is sa regularization parameter. The parameters $\boldsymbol{\theta}_{\text{S}} = [s, b, c_1, c_2, c_3]^T$
provide interpretable control over amplitude ($s$), offset ($b$), and spatial modulations ($c_1,c_2,c_3$).

Learning the optimal correlation structure is cast as a joint optimization problem over both the eigenvalue spectrum and spatial parameters:
\begin{equation}
\mathcal{L}(\{\lambda_k\}, \boldsymbol{\theta}_{\text{S}}) = \left\| \bar{\boldsymbol{\varepsilon}} - \mathbf{C}(\{\lambda_k\}) 
\boldsymbol{\alpha}(\boldsymbol{\theta}_{\text{S}}) \right\|_2^2 + \mathcal{R}(\{\lambda_k\}, \boldsymbol{\theta}_{\text{S}}),
\end{equation}
where $\mathbf{C}(\{\lambda_k\}) = \boldsymbol{\Phi} \text{diag}(\{\lambda_k\}) \boldsymbol{\Phi}^T$ is the correlation matrix with learned eigenvalues, and the regularization term enforces physical constraints:
\begin{equation}
\mathcal{R}(\{\lambda_k\}, \boldsymbol{\theta}_{\text{S}}) = \mu_1 \sum_{k=1}^{Q-1} \max(0, \lambda_{k+1} - \lambda_k) 
+ \mu_2 \sum_{k=1}^{Q} \lambda_k^2 + \mu_3 \|\boldsymbol{\theta}_{\text{GRF}}\|_2^2,
\end{equation}
with $\mu_1, \mu_2, \mu_3 > 0$.

This joint optimization problem is solved using the Tree-structured Parzen Estimator (TPE) algorithm, a sequential model-based optimization (SMBO) approach \cite{mockus1978application,jones1998efficient,bergstra2011algorithms}. The TPE algorithm constructs probabilistic surrogate models of the objective function $\mathcal{L}(\{\lambda_k\}, \boldsymbol{\theta}_{\text{S}})$ and uses an acquisition function to guide the search toward promising regions of the $(Q + 5)$-dimensional parameter space.
TPE builds density models $p(\{\lambda_k\}, \boldsymbol{\theta}_{\text{S}} | \mathcal{L} < \gamma)$ and $p(\{\lambda_k\}, \boldsymbol{\theta}_{\text{S}} | \mathcal{L} \geq \gamma)$ based on previous evaluations, where $\gamma$ represents a quantile threshold. The acquisition function guides the search by maximizing expected improvement, balancing exploration and exploitation across both eigenvalue and spatial parameter dimensions. 
This approach reduces optimization complexity from thousands of individual spatial weights to $(Q + 5)$ parameters, 
while simultaneously learning both the correlation hierarchy and spatial modulations from the observed strain data. 
The learned eigenvalue spectrum captures the intrinsic correlation structure, while the spatial parameters provide physically interpretable control over field variations. 

In this work, we learn $Q=50$ eigenvalues. At each iteration, the optimization proposes a new set of parameters $\boldsymbol{\theta}_{\rm GRF} = [\lambda_1, \dots, \lambda_Q, \, s, b, c_1, c_2, c_3]^T$  which are used to reconstruct the spatial fields.

The reconstruction of the fields proceeds in two separate stages. Firstly, the initial concentration field $c_0(\mathbf{x})$ is inferred by minimizing the discrepancy between the observed strain field and the model prediction at the initial time ($t=0$):
\begin{equation}
\label{eq:obj_func}
\boldsymbol{\theta}_{\rm GRF}^* = \arg \min_{\boldsymbol{\theta}_{\rm GRF} \in \mathbb{R}^n} 
\mathcal{L}(\boldsymbol{\theta}_{\rm GRF}) 
= \int_{\Omega} \left\| 
\boldsymbol{\varepsilon}(\mathbf{x}, t=0; \boldsymbol{\theta}_{\rm GRF}) 
- \boldsymbol{\varepsilon}_{\rm data}(\mathbf{x}, t=0) 
\right\|^2 \, d\mathbf{x}.
\end{equation}
This stage involves only the equilibrium state; thus, the optimization parameter space corresponds solely to $c_0(\mathbf{x})$.

Secondly, the spatial heterogeneity prefactor $k(\mathbf{x})$ is learned during the global optimization that accounts for the full non-equilibrium strain field evolution.

\section*{Supplementary Note 3: Trust-region constrained optimization}
To minimize the objective function defined as:
\begin{equation}
\label{eq:obj_func}
    \mathcal{L}(\boldsymbol{\theta}) = \frac{1}{2} \sum_{i=1}^{N} \int_{\Omega} \left\| \boldsymbol{\varepsilon}(\mathbf{x}, t_i; \boldsymbol{\theta}) - \boldsymbol{\varepsilon}_{\text{data}}(\mathbf{x}, t_i) \right\|^2 \, d\mathbf{x},
\end{equation}

gradient-based optimization is employed in a forward sensitivity-based approach that computes the model sensitivity alongside the model evaluation~\cite{Fletcher2000}. In Eq. \ref{eq:obj_func}, \( N \) represents the number of training snapshots acquired at discrete time steps \( t_i \), \( \boldsymbol{\varepsilon}_{\text{data}} \) is the observed strain field data, and \( \boldsymbol{\varepsilon}(\mathbf{x}, t_i; \boldsymbol{\theta}) \) is the predicted strain field, parameterized by \( \boldsymbol{\theta} \), the parameter vector for all unknowns for the considered inversion case. The optimization approach provides an approximation of the Hessian of the objective function, which becomes increasingly accurate as $\boldsymbol{\theta}$ approaches the true parameters.

\begin{equation}
    \left[\nabla^2 \mathcal{L}\right]_{jk} = \frac{\partial^2 \mathcal{L}}{\partial \boldsymbol{\theta}_j \partial \boldsymbol{\theta}_k} \approx \sum_{i=1}^{N} \int_{\Omega} \frac{\partial \boldsymbol{\varepsilon}}{\partial \boldsymbol{\theta}_j} \cdot \frac{\partial \boldsymbol{\varepsilon}}{\partial \boldsymbol{\theta}_k} \, d\Omega.
\end{equation}

The trust-region constrained optimization algorithm is employed to ensure thermodynamic consistency in solving the inverse problem by enforcing positivity constraints. The general constrained optimization problem is given by:
\begin{equation}
    \min_{\boldsymbol{\theta} \in \mathbb{R}^n} \mathcal{L}(\boldsymbol{\theta}),
    \label{eq:objective}
\end{equation}
subject to the equality and inequality constraints:
\begin{equation}
    c_i(\boldsymbol{\theta}) = 0, \quad i \in \mathcal{E}
    \label{eq:equality_constraints}
\end{equation}
\begin{equation}
    c_j(\boldsymbol{\theta}) \geq 0, \quad j \in \mathcal{I}
    \label{eq:inequality_constraints}
\end{equation}

where $c_i(\boldsymbol{\theta})$ and $c_j(\boldsymbol{\theta})$ are the equality and inequality constraints, respectively. Also, $\mathcal{E}$ and $\mathcal{I}$ are the sets of equality and inequality constraints, respectively. The algorithm approximates the original nonlinear problem using a quadratic model around the current iterate $\boldsymbol{\theta}_i$:
\begin{equation}
    m_i(\mathbf{p}) = \mathcal{L}(\boldsymbol{\theta}_i) + \nabla \mathcal{L}(\boldsymbol{\theta}_i)^T \mathbf{p} + \frac{1}{2} \mathbf{p}^T B_i \mathbf{p}.
    \label{eq:quadratic_model}
\end{equation}

Where $\mathbf{p}$ is the step direction, $B_i$ is an approximation of the Hessian $\nabla^2 \mathcal{L}(\boldsymbol{\theta}_i)$. The step $\mathbf{p}$ is determined by solving the constrained trust-region subproblem:
\begin{equation}
    \min_{\mathbf{p}} m_i(\mathbf{p})
    \label{eq:trust_region_subproblem}
\end{equation}

subject to:
\begin{equation}
    c_i(\boldsymbol{\theta}_i + \mathbf{p}) = 0, \quad i \in \mathcal{E}
    \label{eq:trust_region_eq_constraints}
\end{equation}
\begin{equation}
    c_j(\boldsymbol{\theta}_i + \mathbf{p}) \geq 0, \quad j \in \mathcal{I}
    \label{eq:trust_region_ineq_constraints}
\end{equation}
\begin{equation}
    \|\mathbf{p}\| \leq \Delta_i
    \label{eq:trust_region_radius}
\end{equation}

where $\Delta_i$ is the trust-region radius.
Furthermore, the trust-region radius is updated based on the ratio:
\begin{equation}
    \rho_i = \frac{\mathcal{L}(\boldsymbol{\theta}_i) - \mathcal{L}(\boldsymbol{\theta}_i + \mathbf{p})}{m_i(\mathbf{0}) - m_i(\mathbf{p})}
    \label{eq:ratio_rho}
\end{equation}

The update rules for $\Delta_i$ are:

\begin{equation}
    \Delta_{i+1} =
    \begin{cases}
        \gamma_{\text{inc}} \Delta_i, & \text{if } \rho_i > \eta_2 \\
        \Delta_i, & \text{if } \eta_1 \leq \rho_i \leq \eta_2 \\
        \gamma_{\text{dec}} \Delta_i, & \text{if } \rho_i < \eta_1 
    \end{cases}
    \label{eq:trust_radius_update}
\end{equation}

where $\gamma_{\text{inc}} > 1$ and $0 < \gamma_{\text{dec}} < 1$ are update factors, and $\eta_1, \eta_2$ are predefined thresholds.

Now, the Lagrange function for the constrained problem is expressed as:
\begin{equation}
    L(\boldsymbol{\theta}, \boldsymbol{\lambda}) = \mathcal{L}(\boldsymbol{\theta}) + \sum_{i \in \mathcal{E}} \lambda_i c_i(\boldsymbol{\theta}) + \sum_{j \in \mathcal{I}} \lambda_j c_j(\boldsymbol{\theta}),
    \label{eq:lagrange_function}
\end{equation}
where $\boldsymbol{\lambda}$ are the Lagrange multipliers.
The Hessian approximation $B_i$ is updated using a quasi-Newton formula with projected constraints:
\begin{equation}
    B_{i+1} = B_i + \frac{\mathbf{y}_i \mathbf{y}_i^T}{\mathbf{y}_i^T \mathbf{s}_i} - \frac{B_i \mathbf{s}_i \mathbf{s}_i^T B_i}{\mathbf{s}_i^T B_i \mathbf{s}_i},
    \label{eq:hessian_update}
\end{equation}
where $\mathbf{s}_i = \boldsymbol{\theta}_{i+1} - \boldsymbol{\theta}_i$ and $\mathbf{y}_i = \nabla \mathcal{L}(\boldsymbol{\theta}_{i+1}) - \nabla \mathcal{L}(\boldsymbol{\theta}_i)$ ensure that the update maintains symmetry and positive definiteness.

Finally, the algorithm terminates when the following conditions are met:
\begin{equation}
    \left\|\nabla \mathcal{L}(\boldsymbol{\theta}_i) + \sum_{i \in \mathcal{E}} \lambda_i \nabla c_i(\boldsymbol{\theta}_i) + \sum_{j \in \mathcal{I}} \lambda_j \nabla c_j(\boldsymbol{\theta}_i) \right\| \leq \epsilon_{\text{tol}},
    \label{eq:stopping_criteria}
\end{equation}

where $\epsilon_{\text{tol}}$ is a predefined tolerance set as $10^{-8}$ in this work.


\begin{figure}
\centering
\includegraphics[width=1\textwidth]{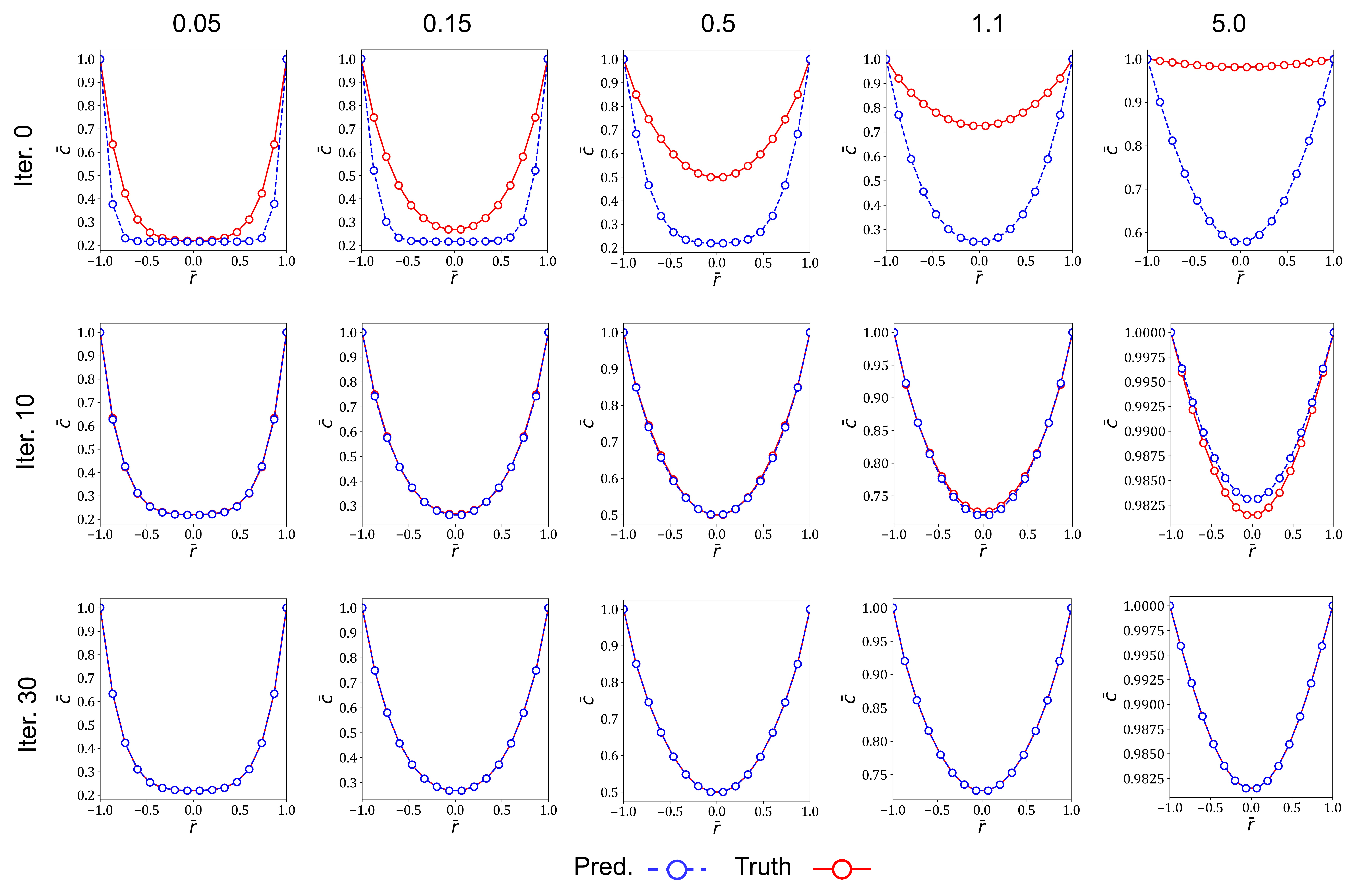}
\caption{\normalsize \textbf{Quantitative convergence of learned diffusivity profiles.}
\\
Radial concentration profiles ($\bar{c}=c/c_{max}$) along the domain diameter (position $\bar{r}=r/R$, where $R$ = $ 10 \, \mu\text{m} $), showing progressive refinement from initial guess to final convergence after 30 iterations.}
\label{fig: Figure S1}
\end{figure}

\begin{figure}
\centering
\includegraphics[width=1\textwidth]{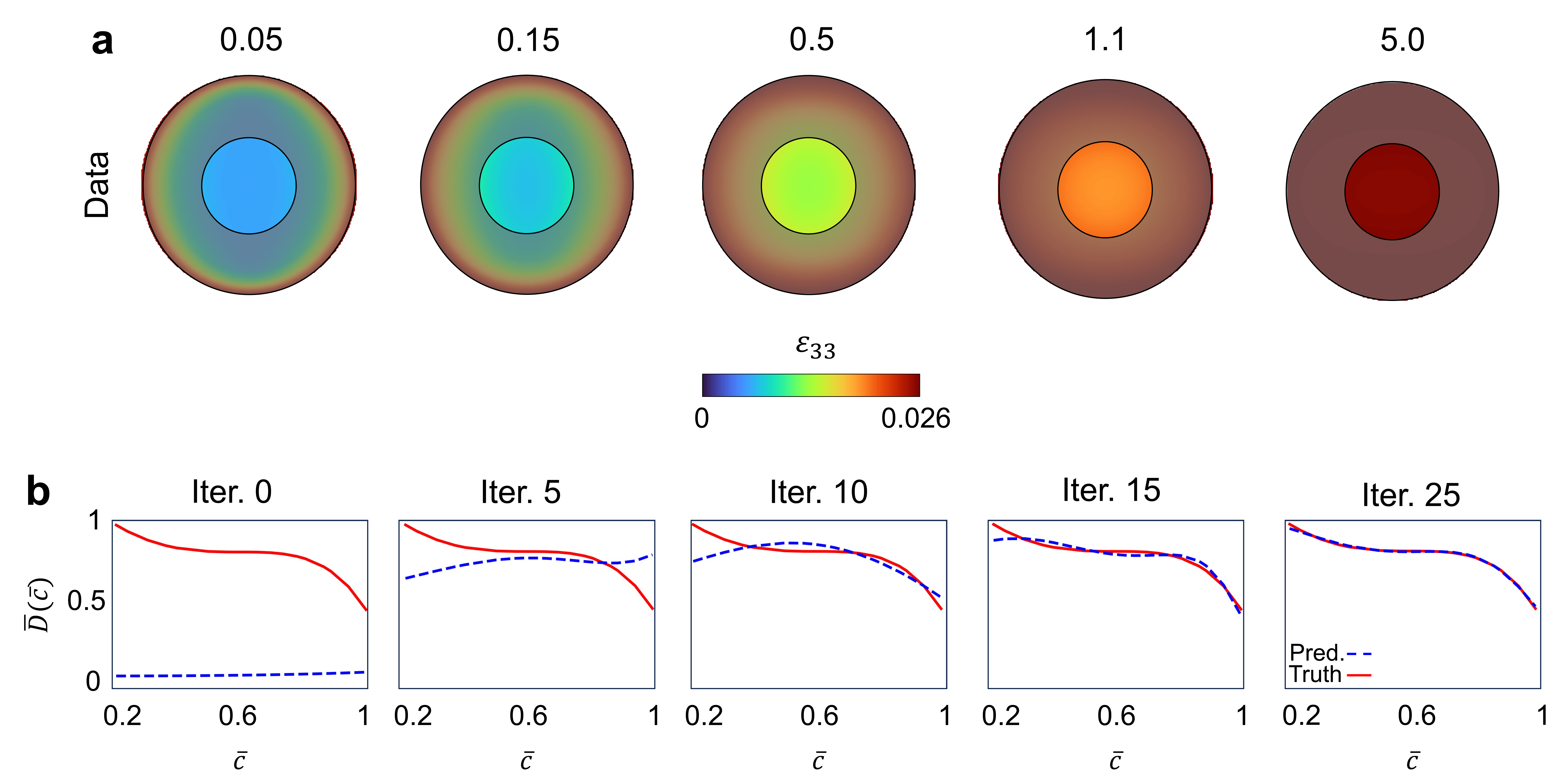}
\caption{\normalsize \textbf{Effect of spatial information loss on inversion of $D(\bar{c})$.} 
\\
\textbf{a} The top row showing five training images
sampled within a larger domain. Numbers above the frames indicate elapsed time (seconds) since the initial frame. \textbf{b} The bottom row
highlighting the evolution of the identified $D(\bar{c})$ from the initial guess to convergence at the 25th iteration.}
\label{fig: Figure S2}
\end{figure}

\begin{figure}
\centering
\includegraphics[width=1\textwidth]{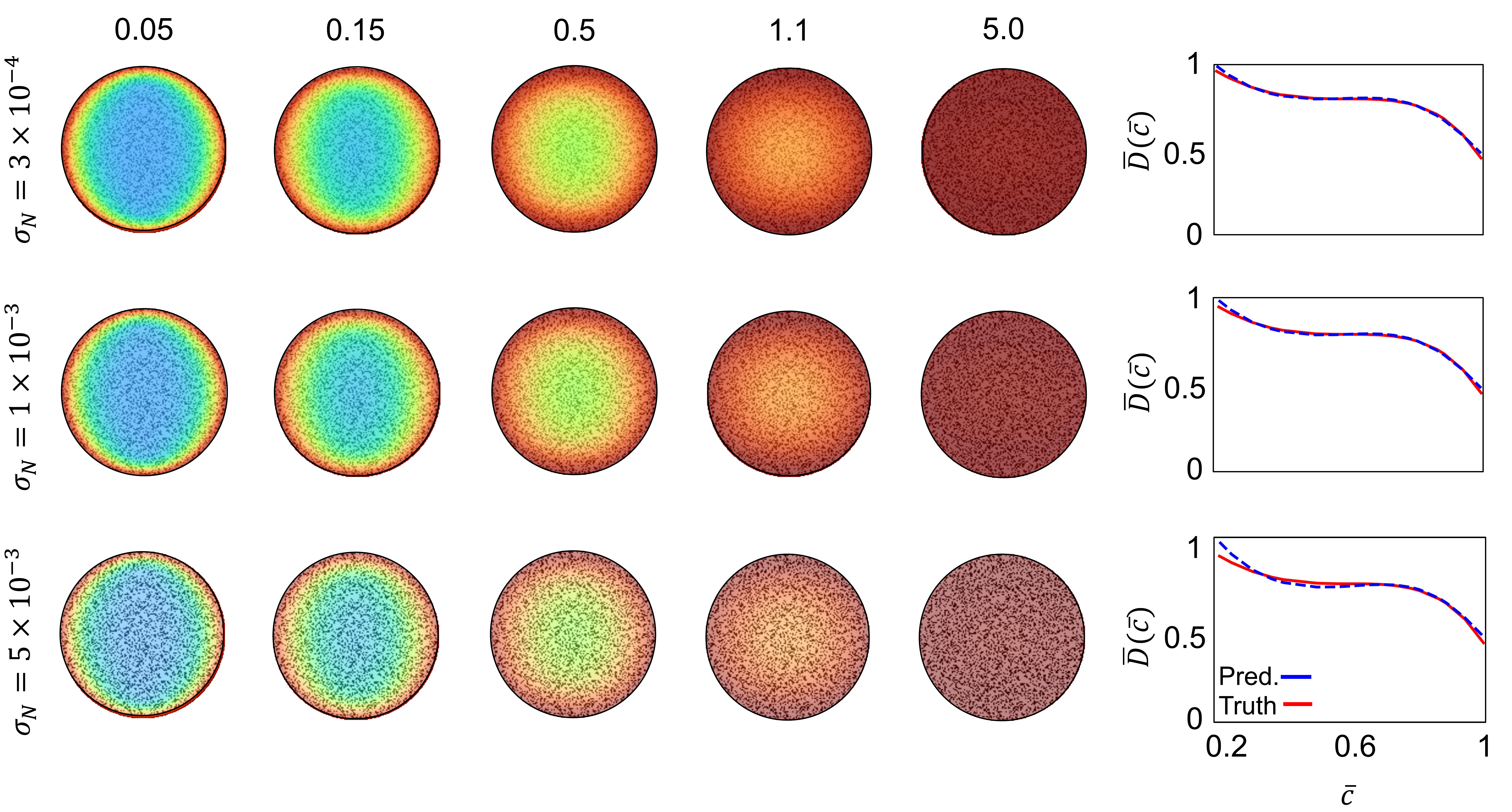}
\caption{\normalsize \textbf{Effect of measurement noise on the inversion of $\bar{D}(\bar{c})$.} \\
Each row corresponds to a different noise level ($\sigma_N=(3 \times 10^{-4},1 \times 10^{-3},5 \times 10^{-3})$ ), showing the input strain images and the inferred diffusivity $\bar{D}(\bar{c})$. The results highlight the framework's robustness to noisy data. However, with an increase in noise amplitude, the agreement between the inferred and true diffusivity slightly deteriorates. Numbers above the frames indicate elapsed time (seconds) since the initial frame.}
\label{fig: Figure S3}
\end{figure}


\begin{figure}
\centering
\includegraphics[width=1\textwidth]{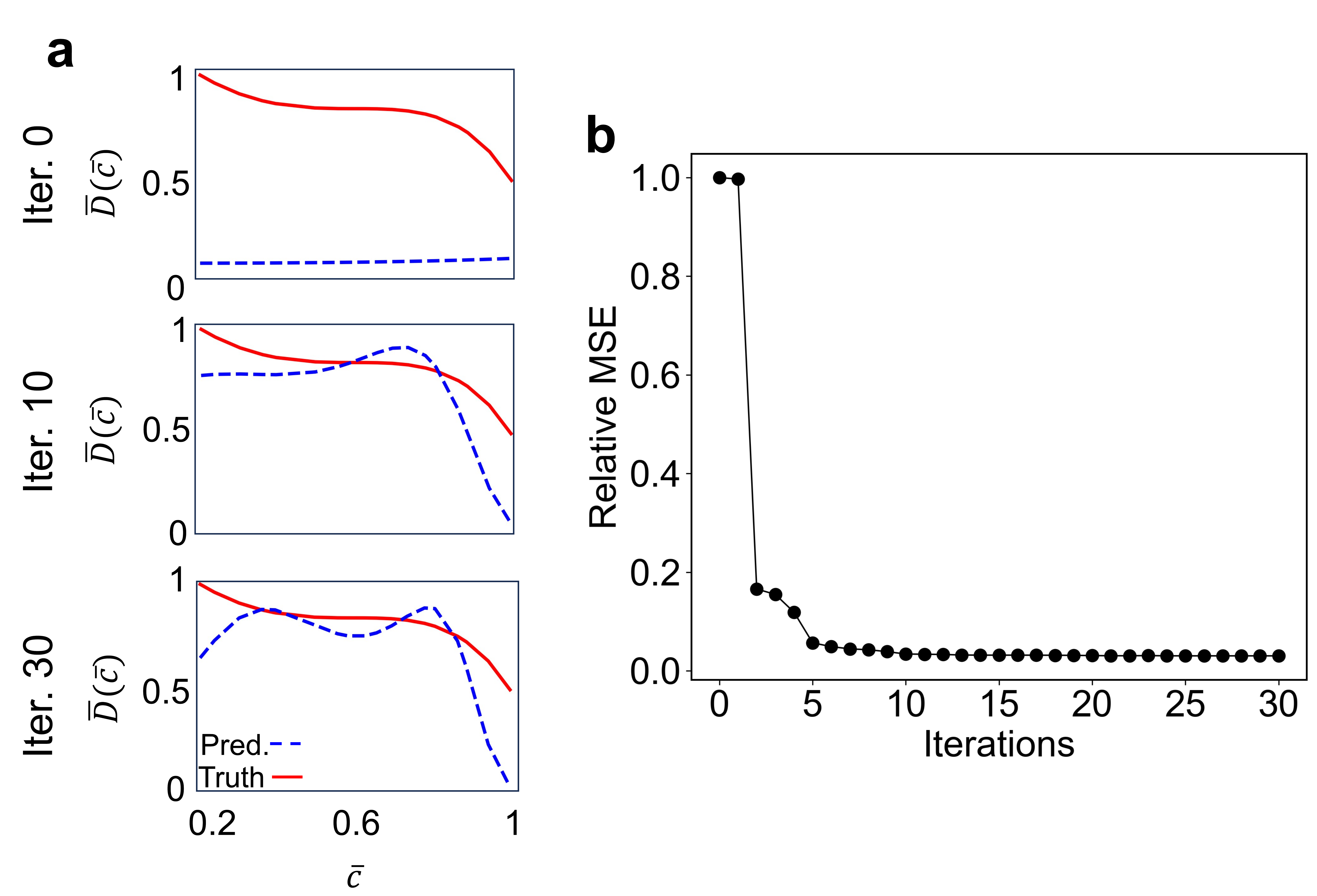}
\caption{\normalsize \textbf{Effect of mechanical information loss on inversion of $\bar{D}(\bar{c})$.} 
\\
A 10\% perturbation is applied to the stiffness constants $C_{11}$ and $C_{33}$ to induce an elevated misfit level that can occur during parameter identification. Five training images of each strain field component are used in the inversion.\\ \textbf{a} The progression of the inversion of $\bar{D}(\bar{c})$ from the initial guess to the final iteration. \textbf{b} The evolution of the relative MSE is shown on the panel on the right.}
\label{fig: Figure S5}
\end{figure}

\begin{figure}
\centering
\includegraphics[width=\textwidth]{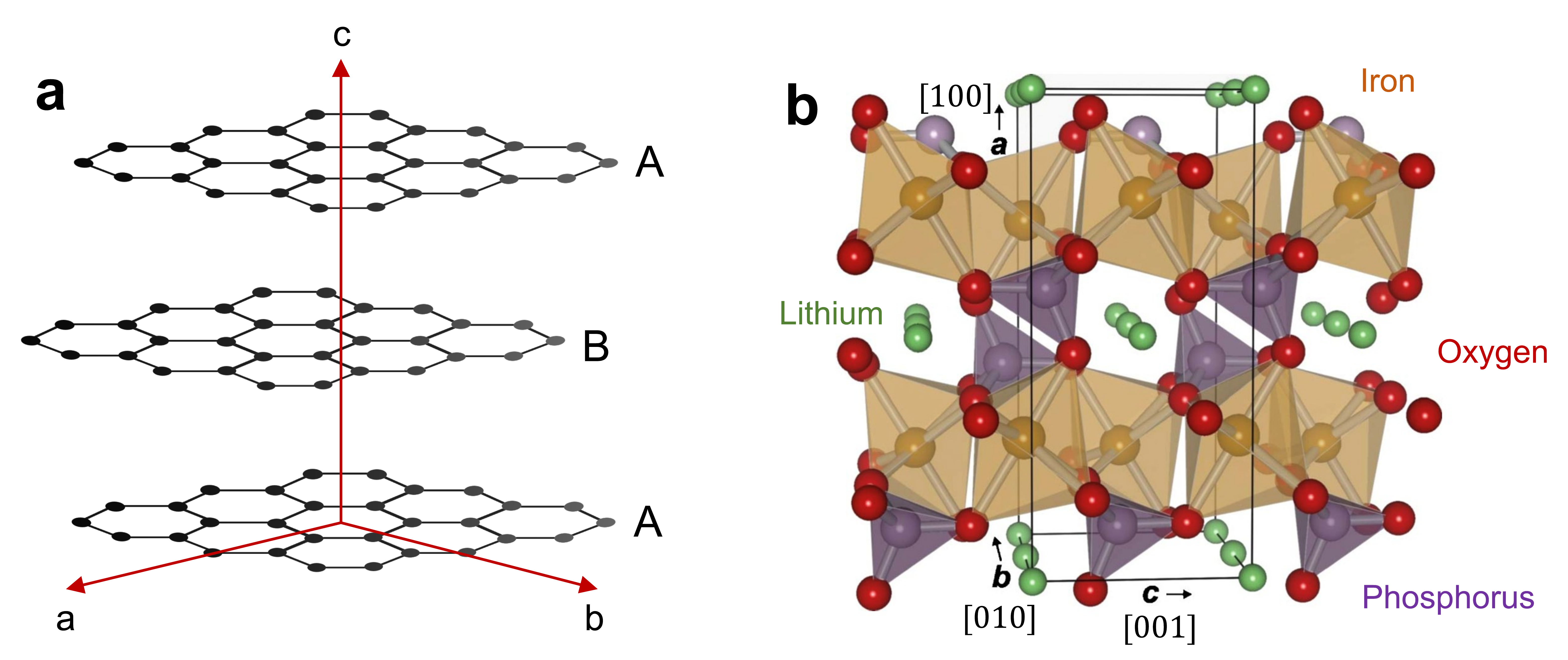}
\caption{\normalsize
\textbf{Crystal lattice structures of the studied electrode systems.} \\
\textbf{a} Graphite, exhibiting a hexagonal layered structure with carbon atoms forming stacked honeycomb sheets. The weak van der Waals bonding between layers governs its anisotropic mechanical and electronic properties. \textbf{b} Olivine LiFePO$_4$, featuring an orthorhombic framework. Lithium ions occupy one-dimensional channels along the [010] direction, enabling facile ionic diffusion. Adapted from Maxisch \textit{et al.} \cite{maxisch2006elastic}.
}
\label{fig: Figure S6}
\end{figure}

\begin{figure}
\centering
\includegraphics[width=1\textwidth]{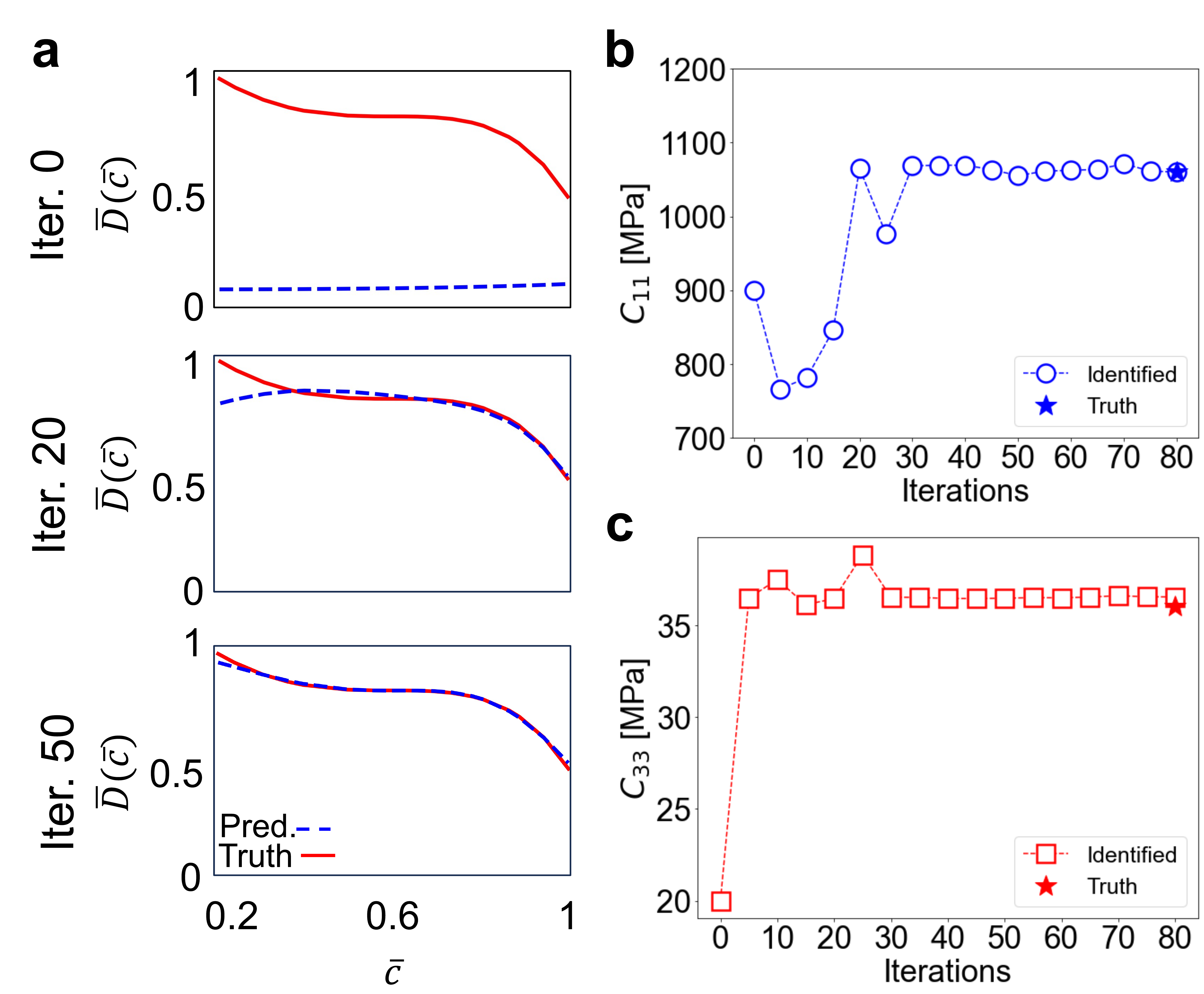}
\caption{\normalsize \textbf{Joint inversion of the concentration-dependent diffusivity $\bar{D}(\bar{c})$ and elastic moduli $C_{11}$ and $C_{33}$.} \\
We simultaneously infer the diffusivity profile $\bar{D}(\bar{c})$ and the principal stiffness components $C_{11}$ and $C_{33}$, corresponding to the dominant elastic responses along the crystallographic $a$- and $c$-axes, respectively. The inversion process converges within 50 iterations, yielding accurate reconstructions of all targeted parameters.\\
\textbf{a} Evolution of $\bar{D}(\bar{c})$ from the initial guess to convergence. \textbf{b} Convergence behavior of the inferred $C_{11}$. \textbf{c} Convergence behavior of the inferred $C_{33}$.}
\label{fig: Figure S7}
\end{figure}

\begin{figure}
\centering
\includegraphics[width=1\textwidth]{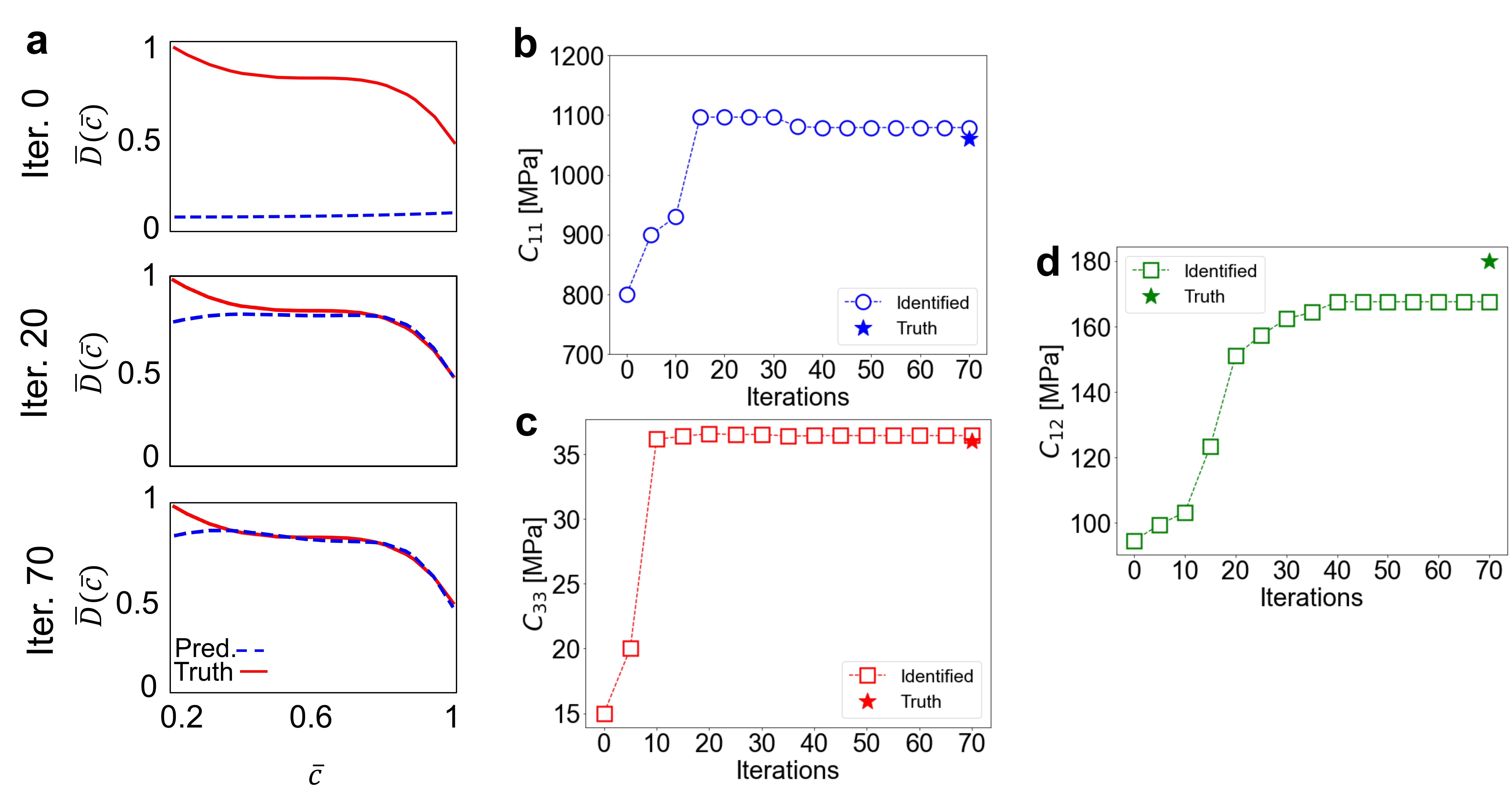}
\caption{\normalsize \textbf{Joint inversion of the concentration-dependent diffusivity $\bar{D}(\bar{c})$ and elastic moduli $C_{11}$, $C_{33}$, and $C_{12}$.} \\
We perform simultaneous inference of the concentration-dependent diffusivity profile $\bar{D}(\bar{c})$, the principal elastic moduli $C_{11}$ and $C_{33}$, governing deformation along the crystallographic $a$- and $c$-axes—and the off-diagonal stiffness component $C_{12}$, which characterizes the coupling between orthogonal in-plane strains. The inversion procedure converges within 70 iterations. While the general trends in $\bar{D}(\bar{c})$, $C_{11}$, and $C_{33}$ are reasonably captured, $C_{12}$ is notably underestimated. \\
\textbf{a} Evolution of $\bar{D}(\bar{c})$ from the initial guess to the final estimate. \textbf{b} Convergence behavior of the inferred $C_{11}$. \textbf{c} Convergence behavior of the inferred $C_{33}$. \textbf{d} Convergence behavior of the inferred $C_{12}$.}
\label{fig: Figure S8}
\end{figure}

\begin{figure}
\centering
\includegraphics[width=1\textwidth]{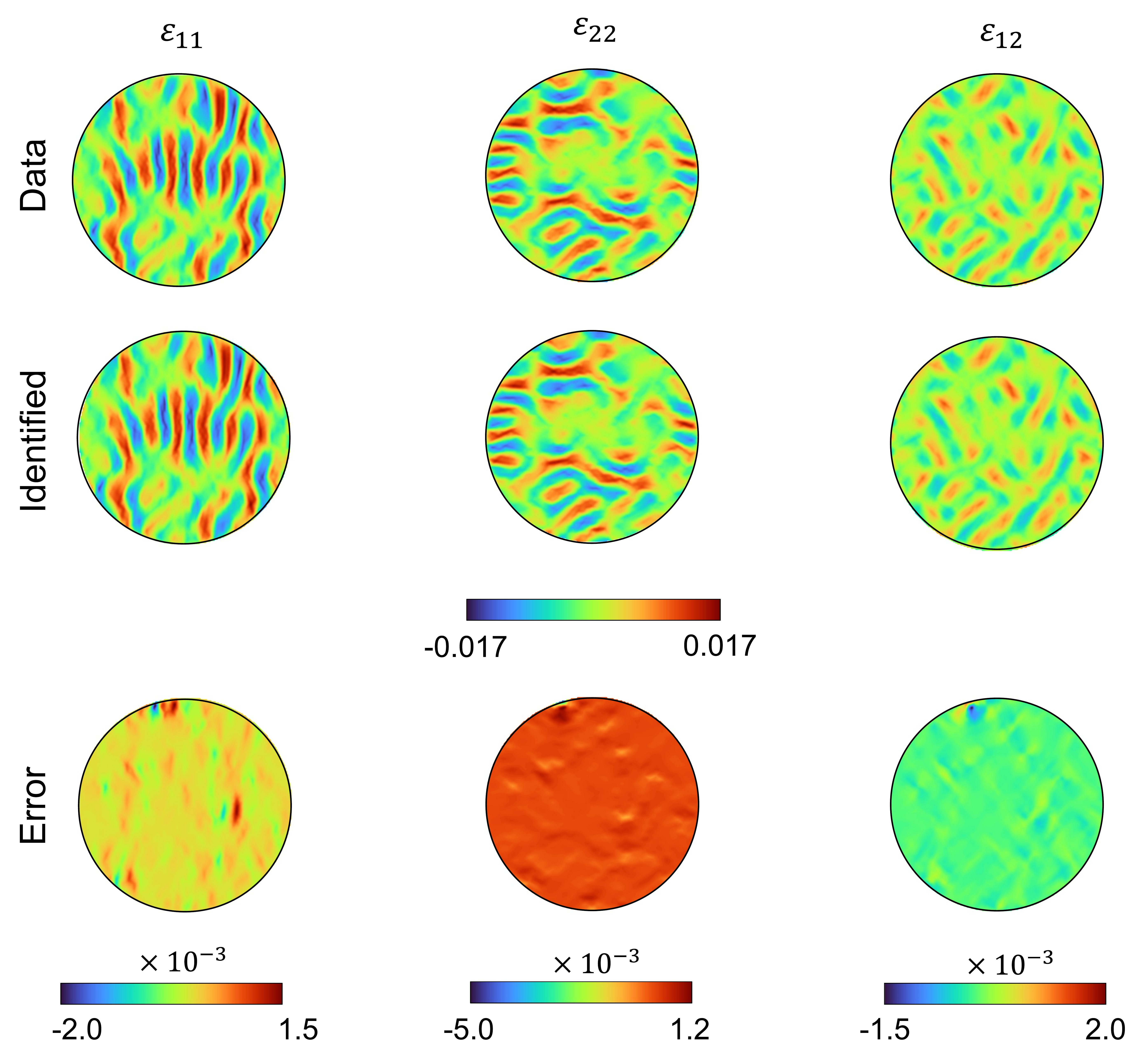}
\caption{\normalsize \textbf{Strain field reconstruction accuracy across additional tensor components for learning pattern formation.}
\\
 Quantitative comparison between ground truth strain fields $\varepsilon_{11}$, $\varepsilon_{22}$, and $\varepsilon_{12}$ and predictions at convergence. Also highlighted is the error plot, showing minimal discrepancies between the ground truth and identified strain fields.}
 \label{fig: Figure S9}
\end{figure}

\begin{figure}
\centering
\includegraphics[width=1\textwidth]{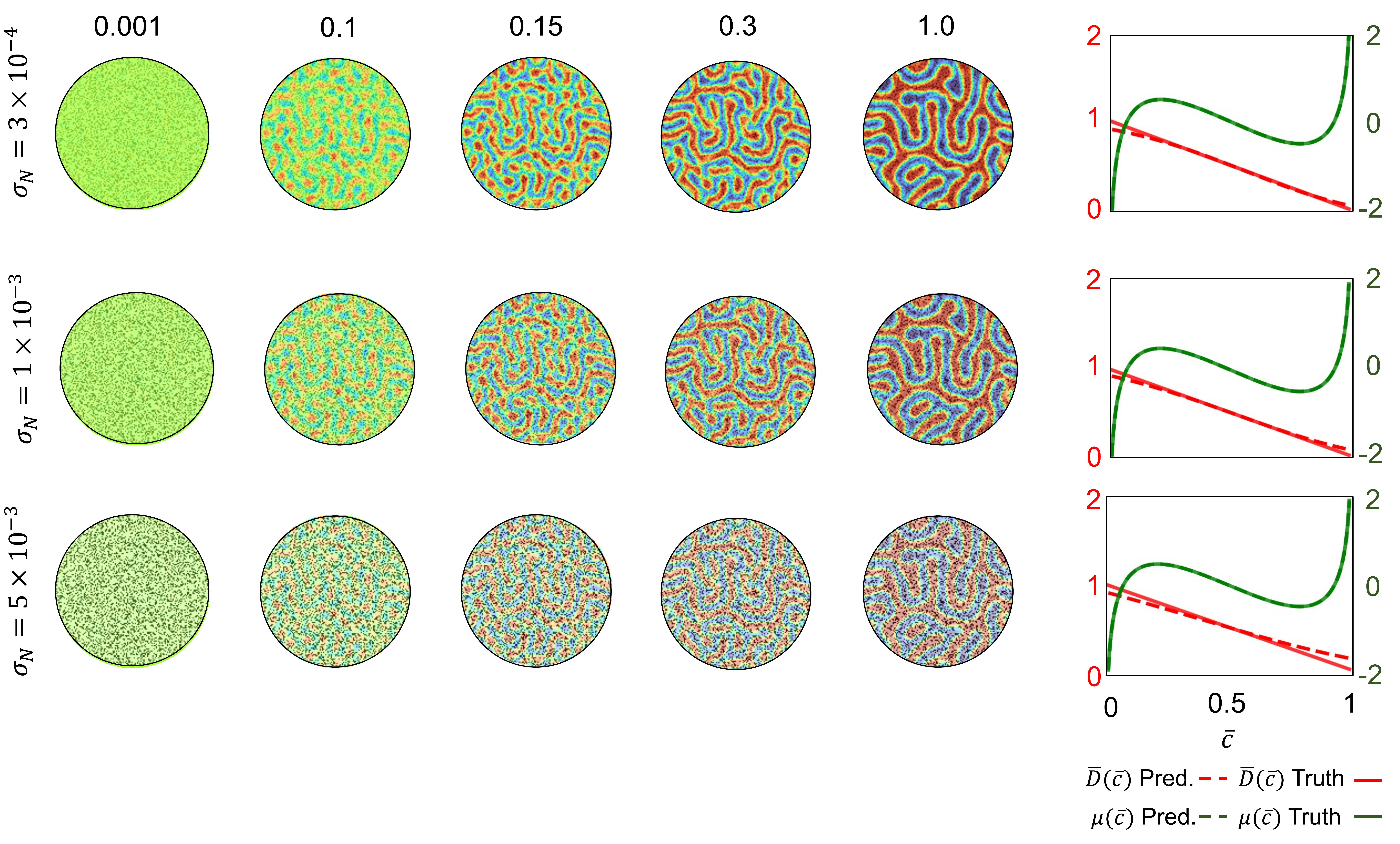}
\caption{\normalsize \textbf{Effect of noise on the robustness of the inversion of $\bar{D}(\bar{c})$ and $\mu (\bar{c})$.}
\\
Each row highlights strain fields at time frames corrupted with noise of increasing Gaussian noise levels  ($\sigma_N=(3 \times 10^{-4},1 \times 10^{-3},5 \times 10^{-3})$ ) of signal amplitude. At the end of each row is the inferred $\bar{D}(\bar{c})$ and $\mu (\bar{c})$ at each noise level. The results highlight the framework's robustness to noisy data. However, with an increase in noise amplitude, the agreement between the inferred and true diffusivity slightly deteriorates.}
\label{fig: Figure S10}
\end{figure}


\begin{figure}
\centering
\includegraphics[width=1\textwidth]{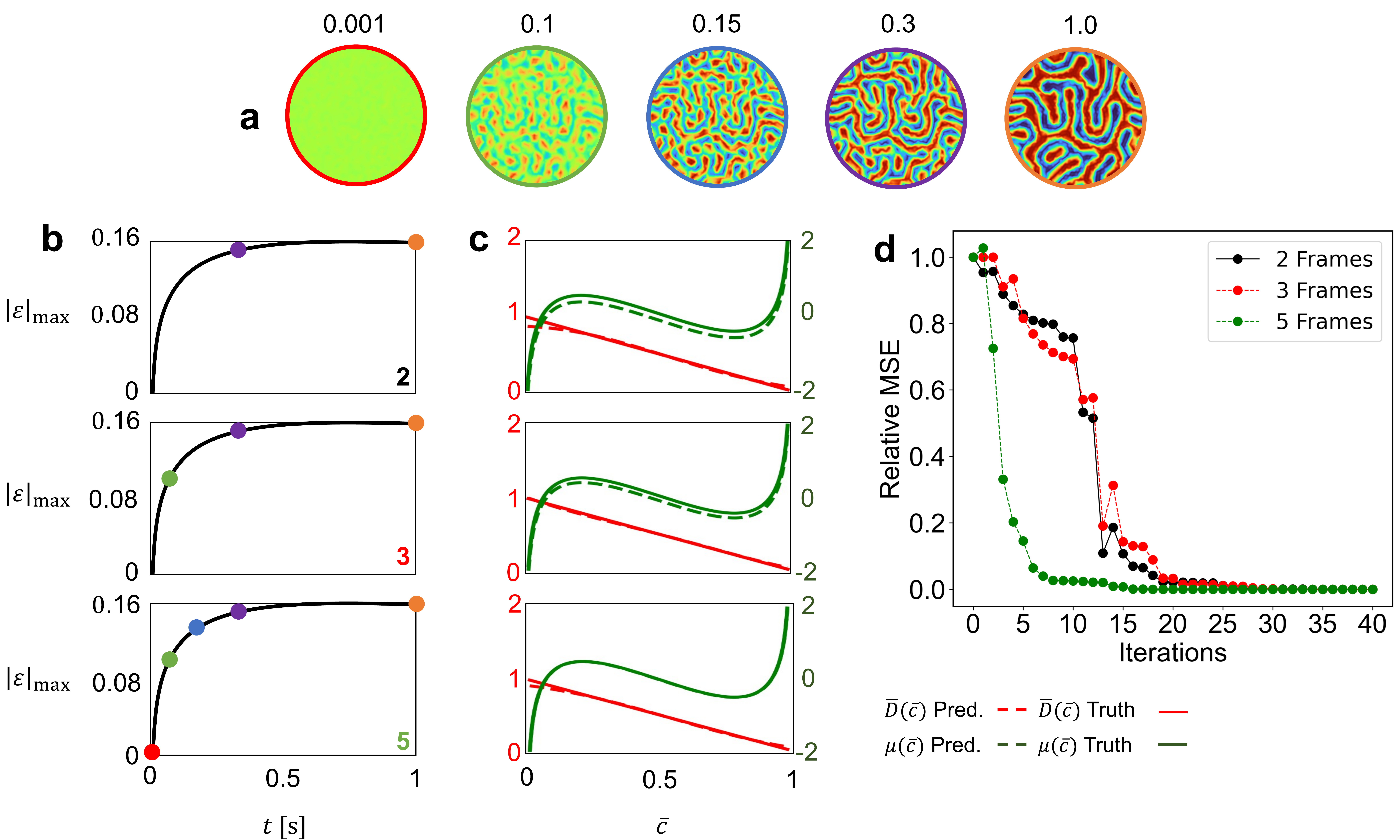}
\caption{\normalsize \textbf{Effect of temporal information loss on inverse learning of pattern formation from strain fields.}
\\
\textbf{a} Candidate strain field snapshots used for case-wise inverse learning. \textbf{b} Corresponding $|\varepsilon|_{\max}
$–$\bar{t}$ trajectories showing the temporal locations of selected training images. \textbf{c} Learned diffusivity $\bar{D}(\bar{c})$ and chemical potential $\mu(\bar{c})$ profiles after convergence. \textbf{d} Evolution of the relative mean squared error (MSE) during training for each case.}
\label{fig: Figure S12}
\end{figure}

\begin{figure}
\centering
\includegraphics[width=1\textwidth]{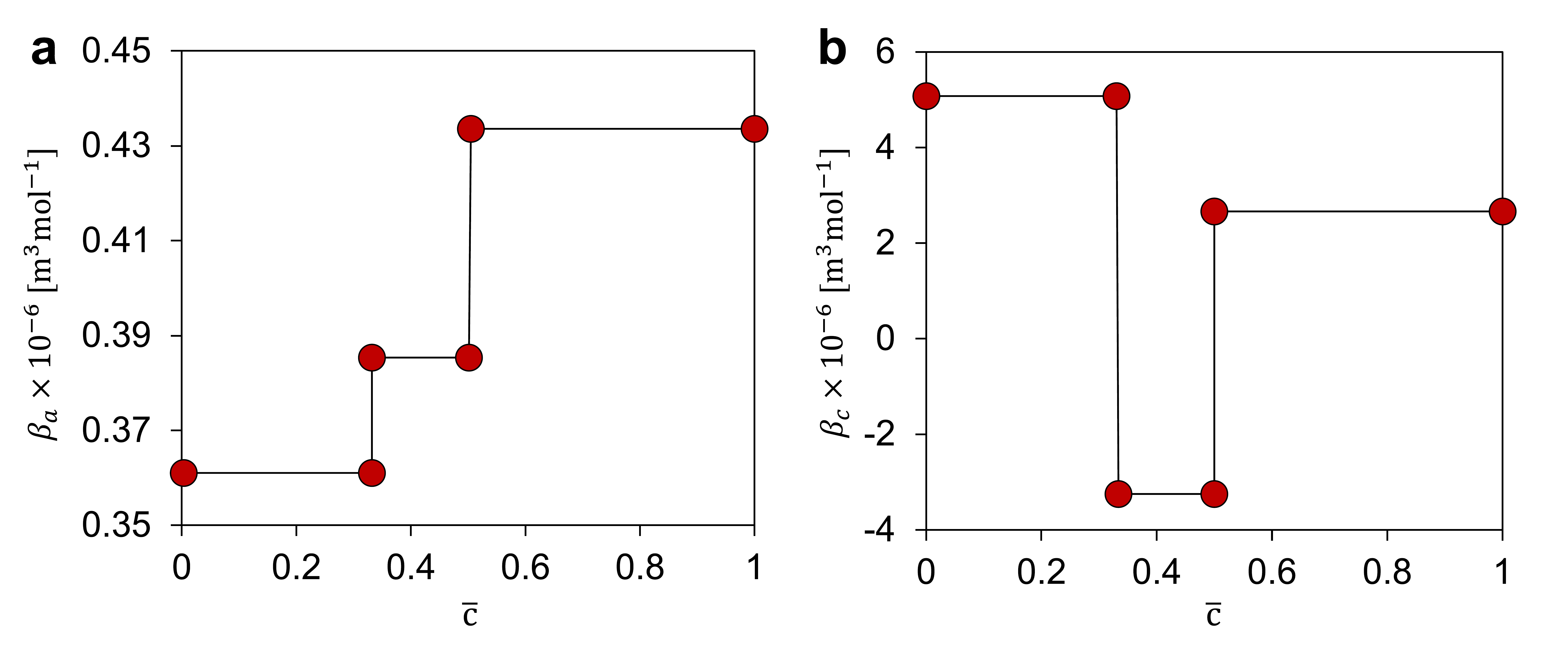}
\caption{\normalsize \textbf{Anisotropic chemical expansion of graphite crystal.} 
\\
\textbf{a} Lattice parameter $a$ expansion coefficient ($\beta_a$) versus normalized lithium concentration ($\bar{c}$). \textbf{b} Lattice parameter $c$ expansion coefficient ($\beta_c$) versus normalized lithium concentration ($\bar{c}$). Adapted from Taghikhani \textit{et al.} \cite{taghikhani2020chemo}.}
\label{fig: Figure S13}
\end{figure}

\begin{figure}
\centering
\includegraphics[width=0.8\textwidth]{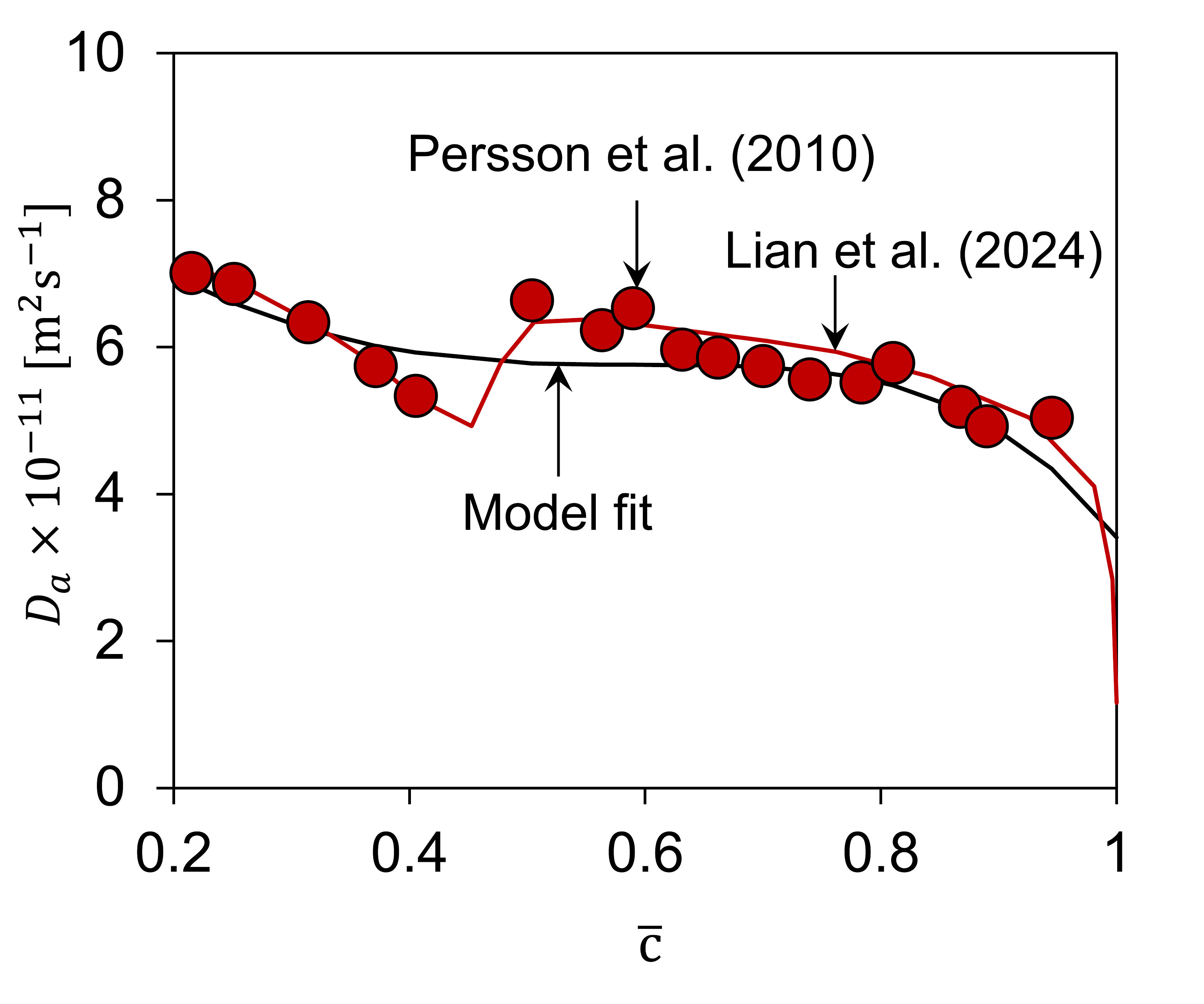}
\caption{\normalsize \textbf{Concentration-dependent lithium diffusivity in graphite.}
\\
Transverse lithium diffusion coefficient ($D_{a}$, m$^2$/s) as a function of normalized lithium concentration ($\bar{c}$). 
Red circles: Experimental data from Persson et al.~\cite{persson2010thermodynamic}. Red solid curve: Interpolation of experimental data by Huada et al \cite{Lian2024}, accounting for phase transitions. Black solid curve: Fitting function used as ground truth in the inverse problem.}
\label{fig: Figure S14}
\end{figure}

\begin{figure}
\centering
\includegraphics[width=0.8\textwidth]{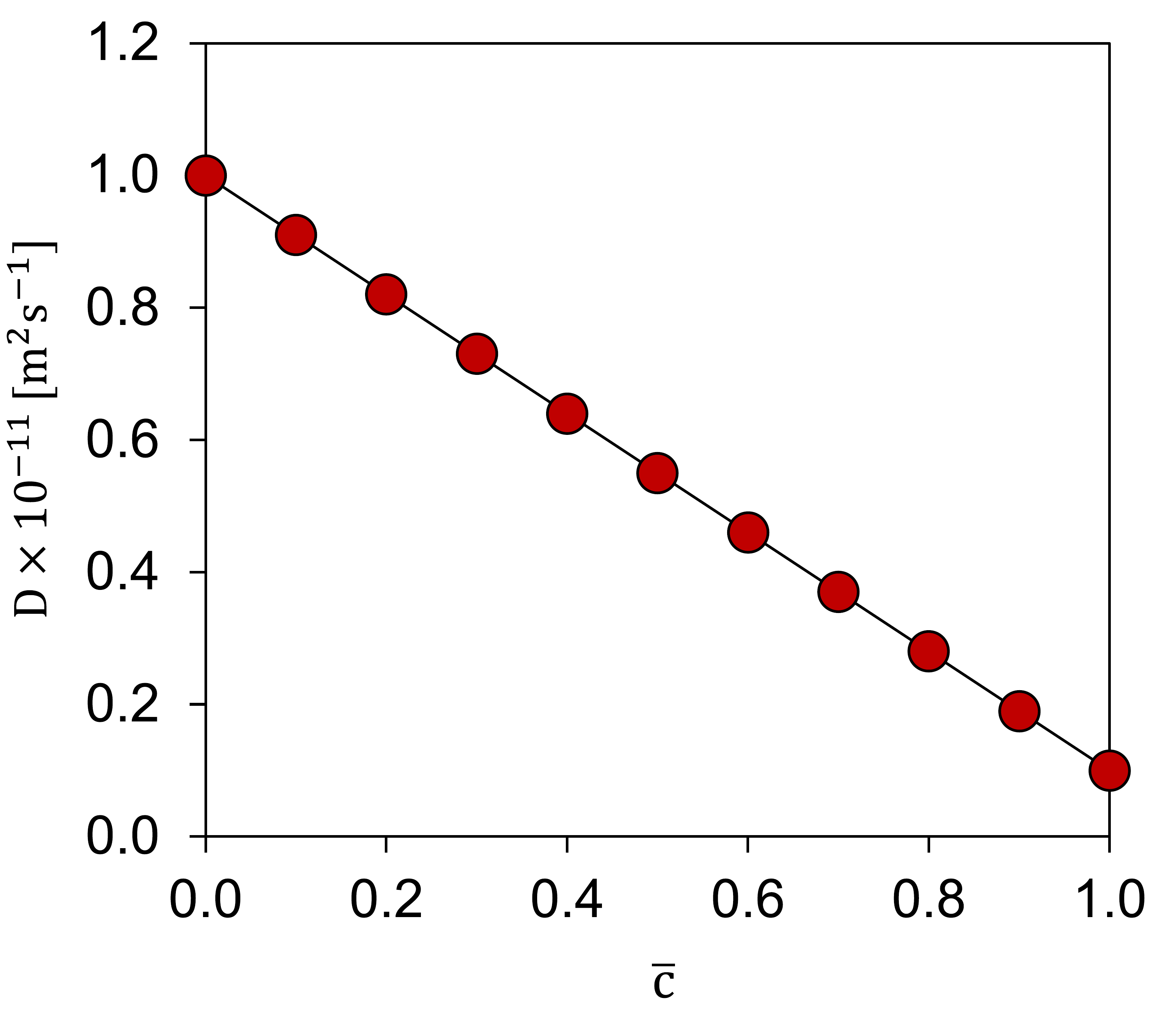}
\caption{\normalsize \textbf{Lithium diffusivity along the [010] direction in LiFePO$_4$.} 
\\
Diffusion coefficient ($D_{Li}$, m$^2$/s) as a function of normalized lithium concentration ($\bar{c}$) in olivine Li$_x$FePO$_4$ at 300 K. Red dots at $\bar{c}=0$ and at $\bar{c}=1$/: First-principles calculations by Morgan et al. \cite{Morgan2003} for dilute limits ($\bar{c}\to0$: $D_0=10^{-11}$ m$^2$/s; $\bar{c}\to1$: $D_1=10^{-12}$ m$^2$/s). Solid black line: Fitting function used as ground truth in the computational model.}
\label{fig: Figure S15}
\end{figure}


\end{document}